\documentclass{article}
\usepackage{geometry}
\usepackage[utf8]{inputenc}
\usepackage{cite}
\usepackage{url}
\usepackage{amssymb,amsmath,amsthm}
\usepackage{bm}
\usepackage{graphicx}
\usepackage{enumerate}
\usepackage{microtype}
\usepackage{color}
\usepackage{hyperref}
\usepackage{multirow}
\usepackage{array}
\newcolumntype{C}[1]{>{\centering\let\newline\\\arraybackslash\hspace{0pt}}m{#1}}

\usepackage[ruled,linesnumbered,vlined]{algorithm2e}
\graphicspath{{figures/}}

\newtheorem{defn}{\noindent $\mathbf{Definition}$}[section]

\newtheorem{theorem}[defn]{$\mathbf{Theorem}$}

\title{Ellipsoidal Density-Equalizing Map for Genus-0 Closed Surfaces}
\author{Zhiyuan Lyu\thanks{Department of Mathematics, The Chinese University of Hong Kong
  ({zhiyuanlyu@cuhk.edu.hk}).}
\and Lok Ming Lui\thanks{Department of Mathematics, The Chinese University of Hong Kong
  ({lmlui@math.cuhk.edu.hk}).}
\and Gary P. T. Choi\thanks{Department of Mathematics, The Chinese University of Hong Kong
  ({ptchoi@cuhk.edu.hk}).}}

\date{}

\begin{document}

\maketitle
\begin{abstract}
Surface parameterization is a fundamental task in geometry processing and plays an important role in many science and engineering applications. In recent years, the density-equalizing map, a shape deformation technique based on the physical principle of density diffusion, has been utilized for the parameterization of simply connected and multiply connected open surfaces. More recently, a spherical density-equalizing mapping method has been developed for the parameterization of genus-0 closed surfaces. However, for genus-0 closed surfaces with extreme geometry, using a spherical domain for the parameterization may induce large geometric distortion. In this work, we develop a novel method for computing density-equalizing maps of genus-0 closed surfaces onto an ellipsoidal domain. This allows us to achieve ellipsoidal area-preserving parameterizations and ellipsoidal parameterizations with controlled area change. We further propose an energy minimization approach that combines density-equalizing maps and quasi-conformal maps, which allows us to produce ellipsoidal density-equalizing quasi-conformal maps for achieving a balance between density-equalization and quasi-conformality. Using our proposed methods, we can significantly improve the performance of surface remeshing for genus-0 closed surfaces. Experimental results on a large variety of genus-0 closed surfaces are presented to demonstrate the effectiveness of our proposed methods. 
\end{abstract}

\section{Introduction}

Surface parameterization is the process of mapping a complicated surface onto a simpler domain. It is closely related to many tasks in geometry processing including surface registration, texture mapping, surface remeshing, and shape analysis. In recent decades, numerous efforts have been devoted to the development of efficient surface parameterization algorithms with different desired properties~\cite{floater2005surface,sheffer2007mesh,hormann2008mesh,choi2023recent} for various science and engineering applications. 

As genus-0 closed surfaces are topologically equivalent to a sphere, most prior works have focused on the problem of parameterizing genus-0 closed surfaces onto the unit sphere. For conformal parameterization, a vast number of spherical conformal parameterization methods have been developed over the past few decades based on linearization~\cite{haker2000conformal}, harmonic energy minimization~\cite{gu2004genus}, Ricci flow~\cite{chen2013ricci}, conformal curvature flow~\cite{crane2013robust}, quasi-conformal theory~\cite{choi2015flash}, partial welding~\cite{choi2020parallelizable} and Dirichlet energy minimization~\cite{liao2024convergence}. For area-preserving parameterization, the theory and computation of optimal mass transportation (OMT) have been extensively studied~\cite{su2013area,gu2016variational}. Based on OMT, some area-preserving parameterization methods for genus-0 closed surfaces have been developed~\cite{cui2019spherical,pumarola20193dpeople,giri2021open,choi2022adaptive}. Besides the above-mentioned conformal and area-preserving methods, some other prior works have also developed spherical parameterization algorithms with different mapping criteria~\cite{gotsman2003fundamentals,lui2007landmark,lefevre2015spherical,nadeem2016spherical,wang2016bijective,choi2016fast,hu2017advanced,wang2018novel,yueh2021projected,lyu2023two,huang2024fundamental}. While there is a topological equivalence between genus-0 closed surfaces and the sphere, the geometric difference between them may be large. For instance, mapping a genus-0 closed surface with an elongated shape to a sphere may induce large geometric distortion. Therefore, a few recent studies have focused on ellipsoidal parameterizations~\cite{lin2023ellipsoidal,choi2024fast,shaqfa2024spheroidal}, in which an ellipsoid is used as the parameter domain for parameterizing genus-0 closed surfaces. 

Density-equalizing maps~\cite{gastner2004diffusion} are a class of mappings with shape deformation produced based on a prescribed density distribution. More specifically, given a planar domain with a density distribution prescribed in it, a density-equalizing map will deform the domain with the high-density regions enlarged and the low-density regions shrunk. This concept is naturally related to the problem of cartogram creation, which focuses on creating shape deformations of geographical maps to visualize specific sociological data. In classical cartogram methods~\cite{tobler1973continuous}, continuous transformations balancing area targets with conformality were considered for districting. Over the past few decades, density-equalizing maps have been widely used in cartogram creation and data visualization~\cite{dorling2008atlas}. Also, different variants and improved algorithms have been developed~\cite{gastner2005spatial,gastner2018fast,li2018diffusion}. Choi and Rycroft~\cite{choi2018density} developed a novel method for computing surface density-equalizing maps and demonstrated its effectiveness in surface parameterization for simply connected open surfaces. Later, the method was extended to other mapping problems for simply connected open surfaces~\cite{choi2020area,shaqfa2024disk,choi2024hemispheroidal} and volumetric domains~\cite{choi2021volumetric} with applications to medical visualization and shape modeling. Lyu et al.~\cite{lyu2024bijective} developed a surface parameterization method for multiply connected open surfaces by combining density-equalizing maps and quasi-conformal maps. More recently, they have developed spherical density-equalizing mapping methods for genus-0 closed surfaces~\cite{lyu2024spherical}. Using their methods, spherical area-preserving parameterizations and spherical parameterizations with controlled area change can be achieved. However, as mentioned above, mapping genus-0 closed surfaces with extreme geometry onto a spherical domain may induce large geometric distortion. For instance, the spherical area-preserving parameterization of an elongated surface may possess low area distortion but extremely high angular distortion, which is undesirable for many practical applications. Moreover, ellipsoidal models play an important role in geographic reference systems (e.g. WGS 84), and ellipsoidal mappings can be naturally utilized in cartograms and simulations of the Earth or other celestial bodies.

\begin{figure}[t!]
    \centering
    \includegraphics[width=\textwidth]{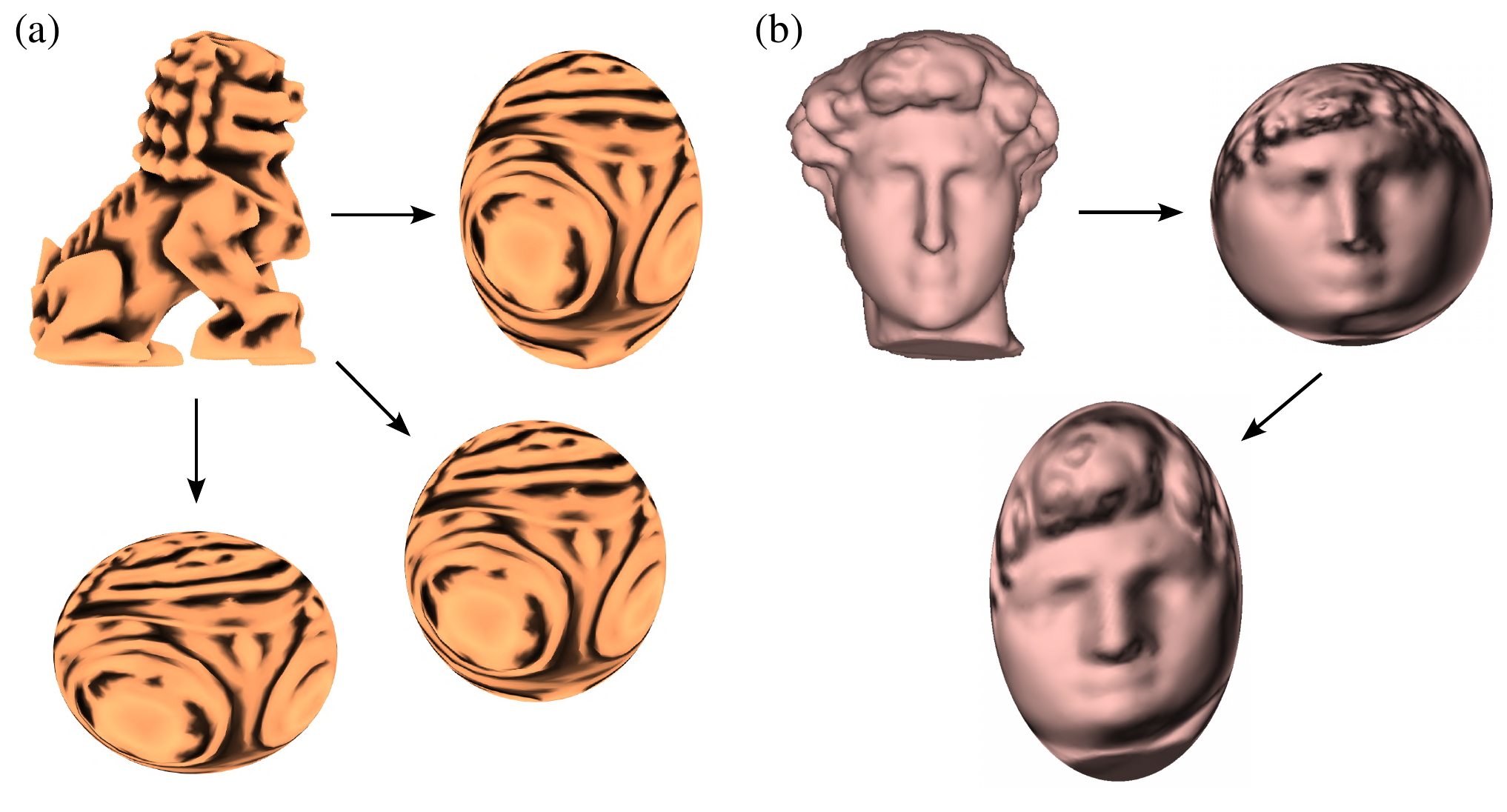}
    \caption{\textbf{An illustration of the proposed ellipsoidal density-equalizing mapping method (EDEM) and ellipsoidal density-equalizing quasi-conformal mapping method (EDEQ).} (a)~Given any input genus-0 closed surface, we can apply the proposed EDEM method to compute ellipsoidal density-equalizing maps to ellipsoidal domains with different prescribed elliptic radii. (b)~Given any input genus-0 closed surface (top left), we can start with an initial spherical parameterization (top right) and then apply the proposed EDEQ method to simultaneously optimize the elliptic radii of the domain and the mapping onto it, thereby achieving an ellipsoidal density-equalizing quasi-conformal map with minimal geometric distortions (bottom).}
    \label{fig:front_add}
\end{figure}

Motivated by the above works, here we develop a novel method for computing ellipsoidal density-equalizing maps for genus-0 closed surfaces in $\mathbb R^3$. Specifically, given a genus-0 closed surface, we compute density-equalizing maps onto ellipsoidal domains with different prescribed radii based on a density distribution encoding the desired mapping effect (Fig.~\ref{fig:front_add}(a)). This allows us to easily achieve ellipsoidal area-preserving parameterizations and ellipsoidal parameterizations with controlled area change. We then further propose an algorithm for computing ellipsoidal density-equalizing quasi-conformal maps, which combines density-equalizing maps and quasi-conformal maps to achieve a balance between area and angle distortions. Moreover, throughout the algorithm, we can optimize the shape of the target ellipsoidal domain to further reduce overall geometric distortion (Fig.~\ref{fig:front_add}(b)). This provides us with an effective and automatic way of representing any genus-0 closed surface by an optimal ellipsoid. We apply our proposed methods for surface remeshing of genus-0 closed surfaces and demonstrate the improvement over prior approaches via various examples. 

The organization of this paper is as follows. In Section~\ref{sec:background}, we introduce the mathematical background of our work. In Section~\ref{sec:main}, we describe our proposed methods for computing ellipsoidal density-equalizing maps (EDEM) and ellipsoidal density-equalizing quasi-conformal maps (EDEQ). In Section~\ref{sect:experiment}, we present experimental results of our proposed methods on various genus-0 closed surfaces. In Section~\ref{sec:applications}, we describe the application of our proposed methods to genus-0 surface remeshing. Finally, in Section~\ref{sect:discussion}, we conclude this paper and discuss possible future works.

\section{Mathematical background}\label{sec:background}
Our proposed methods are primarily based on the theory of density-equalizing maps and quasi-conformal geometry. The relevant concepts are introduced in this section. 

\subsection{Density-equalizing maps}
\begin{figure}[t]
   \centering
\includegraphics[width=0.9\textwidth]{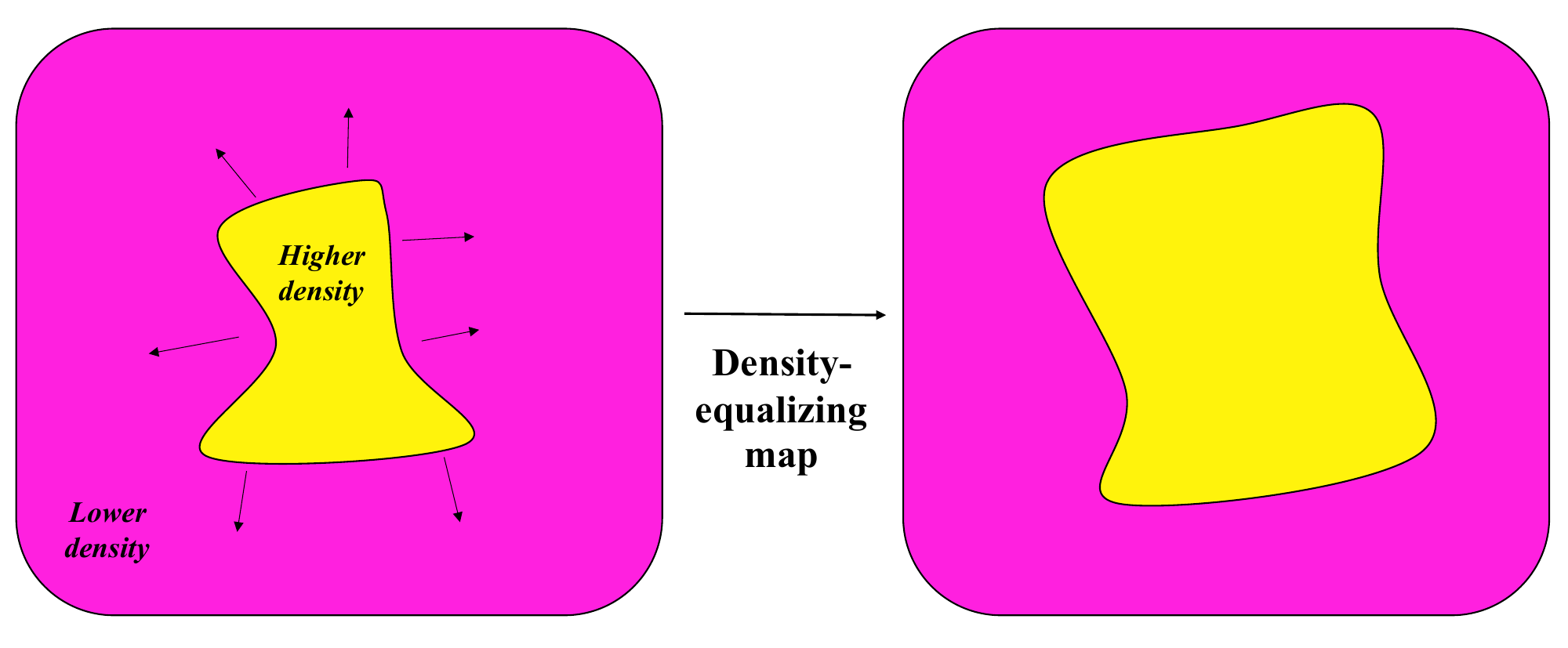}
    \caption{\textbf{An illustration of density-equalizing maps.} The density flow enlarges the regions with high density and shrinks the regions with low density. }
    \label{fig:DEM_illustration}
\end{figure}

Gastner and Newmann~\cite{gastner2004diffusion} proposed a method for computing the \emph{density-equalizing maps} for 2D planar domains based on the principle of density diffusion. Given a positive density function $\rho$ defined on the planar domain, the method produces shape deformation following the density gradient. More specifically, note that the advection equation is given by
\begin{equation}\label{advection}
    \frac{\partial \rho}{\partial t} = - \nabla \cdot \mathbf{j},
\end{equation}
where $\mathbf{j} = - \nabla \rho$ is the density flux by Fick's law. This gives the diffusion equation
\begin{equation}\label{diffusion-eq}
    \frac{\partial \rho}{\partial t} = \Delta \rho.
\end{equation}
Now, since the flux $\mathbf{j}$ can be expressed as $\mathbf{j} = \rho \mathbf{v}$, where $\mathbf{v}$ is the velocity field, we have
\begin{equation}
    \mathbf{v} = \frac{\mathbf{j}}{\rho} = -\frac{\nabla \rho}{\rho}.
\end{equation}
Therefore, the position of any tracer particle $\mathbf{r}$ at time $t$ can be traced by:
\begin{equation}\label{displacement}
    \mathbf{r}(t) = \mathbf{r}(0) + \int^{t}_{0} \mathbf{v}(\mathbf{r},\tau) \mathrm{d\tau}.
\end{equation}
where $\mathbf{r}(0)$ is the initial position. Taking the time $t\to \infty$, the density $\rho$ will be fully equalized on the entire domain, and the resulting shape deformation is a density-equalizing map. In particular, as the shape deformation is induced by the density diffusion process, it is easy to see that high-density regions will be enlarged throughout the process and low-density regions will be shrunk (see Fig.~\ref{fig:DEM_illustration} for an illustration).

The above density-equalizing mapping method has been widely applied to cartogram creation and sociological data visualization. Specifically, the input density $\rho$ is commonly defined as some prescribed quantity (known as the ``population'') per unit area, where the ``population'' can either be the actual population of a certain region on the geographical map, the income of a region, or any other sociological data to be visualized. 

In recent years, the density-equalizing mapping method has been further introduced to the field of surface parameterization. In particular, Choi et al.~\cite{choi2018density,choi2020area} proposed methods for computing density-equalizing maps for simply connected open surfaces in $\mathbb{R}^3$. Lyu et al.~\cite{lyu2024bijective} developed parameterization methods for multiply connected open surfaces based on density-equalizing maps. More recently, Lyu et al.~\cite{lyu2024spherical} developed a method for computing spherical density-equalizing maps for genus-0 closed surfaces.

\subsection{Quasi-conformal theory}
\begin{figure}[t]
   \centering
   \includegraphics[width=0.9\textwidth]{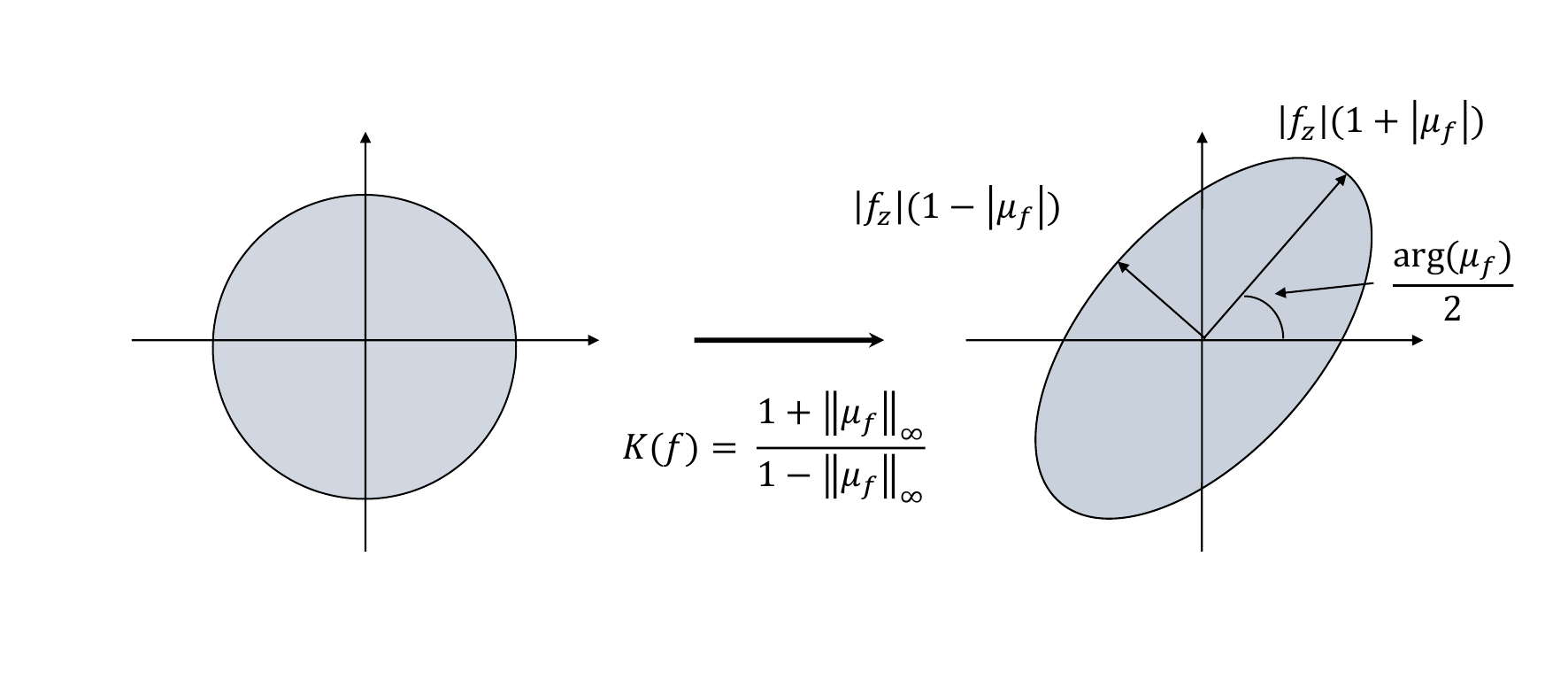}
    \caption{\textbf{An illustration of quasi-conformal maps.} The Beltrami coefficient determines the conformality distortions }
    \label{fig:QC_illustration}
\end{figure}

It is well-known that conformal maps~\cite{nehari2012conformal} are angle-preserving. Intuitively, they map infinitesimal circles to infinitesimal circles. \emph{Quasi-conformal maps}~\cite{lehto1973quasiconformal,ahlfors2006lectures} are a generalization of conformal maps taking infinitesimal circles to infinitesimal ellipses with bounded eccentricity. Mathematically, a quasi-conformal map $f: \overline{\mathbb C} \rightarrow \overline{\mathbb C}$ satisfies the Beltrami equation:
\begin{equation}\label{eq:Beltrami_eq}
    \frac{\partial f}{\partial \Bar{z}} = \mu(z) \frac{\partial f}{\partial z}
\end{equation}
for some complex-valued function $\mu$ satisfying $\|\mu \|_{\infty} < 1$. Here, $\mu$ is called the \emph{Beltrami coefficient}, which encodes important information about the conformal distortion of the map $f$. Specifically, $f$ is a conformal map if and only if $\mu = 0$, as Eq.~\eqref{eq:Beltrami_eq} becomes the Cauchy--Riemann equation. Intuitively, around a point $z_0 \in \mathbb C$, the first order approximation of $f$ can be expressed as:
\begin{equation}\label{eqt:first_order_approximation}
    f(z) \approx f(z_0) + f_{z}(z_0)(z-z_0) + f_{\Bar{z}}(z_0)\overline{(z-z_0)} = f(z_0) + f_{z}(z_0)(z-z_0 + \mu(z_0)\overline{(z-z_0)}).
\end{equation}
The above approximation suggests that $f$ maps an infinitesimal circle centered at $z_0$ to an infinitesimal ellipse centered at $f(z_0)$. Additionally, we can determine the angles and scaling factors of the maximal magnification and maximal shrinkage using $\mu$ (see Fig.~\ref{fig:QC_illustration}). More specifically, the angle of maximal magnification is given by $\operatorname{arg}(\mu(z_0))/2$, with the magnifying factor $|f_{z}(z_0)|(1+|\mu(z_0)|)$. Also, the angle of maximal shrinkage is given by $(\operatorname{arg}(\mu(z_0))+\pi)/2$, with the shrinking factor $|f_{z}(z_0)|(1-|\mu(z_0)|)$. The maximal dilation of $f$ is $K(f) = \frac{1+\|\mu \|_{\infty}}{1-\|\mu\|_{\infty}}$. 

Besides encoding important geometric information about the associated quasi-conformal map, the Beltrami coefficient is also related to the bijectivity of the mapping~\cite{lehto1973quasiconformal,ahlfors2006lectures}:
\begin{theorem}\label{qc_bijective}
    If $f$ is a $C^1$ mapping satisfying $\|\mu_f \|_{\infty}<1$, then $f$ is bijective. 
\end{theorem}

Moreover, the Beltrami coefficient of a composition of two quasi-conformal maps can be expressed in the following way. Let $f,g: \overline{\mathbb{C}} \rightarrow \overline{\mathbb{C}}$ be two quasi-conformal maps with Beltrami coefficient $\mu_f$ and $\mu_g$, respectively. The Beltrami coefficient of the composition map $g\circ f$ is given by
\begin{equation}\label{composition_BC}
    \mu_{g \circ f} = \dfrac{\mu_f + (\mu_{g}\circ f) \tau }{1 + \overline{\mu_f}(\mu_g \circ f) \tau},
\end{equation}
where $\tau = \overline{f_z} / f_z$. In particular, if $g$ is a conformal map, then $\mu_{g \circ f} = \mu_f$. In other words, composing a conformal map with a given quasi-conformal map will not change its Beltrami coefficient.

Lui et al.~\cite{lui2013texture} proposed an algorithm called \emph{Linear Beltrami Solver (LBS)} for efficiently reconstructing a quasi-conformal mapping based on a given Beltrami coefficient $\mu$. Specifically, for any given $\mu = \xi + i\tau$, the corresponding quasi-conformal map $f = u + iv$ can be computed by solving the following equations:
\begin{equation} \label{eqt:lbs}
\left\{\begin{array}{l}
\nabla \cdot(A \nabla u)=0 \\
\nabla \cdot(A \nabla v)=0
\end{array}\right.
\end{equation}
where 
$
A=\left(\begin{array}{cc}
\alpha_1 & \alpha_2 \\
\alpha_2 & \alpha_3
\end{array}\right)
$
with 
\begin{equation}
    \alpha_1=\frac{(\xi-1)^2+\tau^2}{1-\xi^2-\tau^2},  \ \ \alpha_2=-\frac{2 \tau}{1-\xi^2-\tau^2},  \ \ \alpha_3=\frac{(\xi+1)^2+\tau^2}{1-\xi^2-\tau^2}.
\end{equation}
In the discrete case, Eq.~\eqref{eqt:lbs} becomes two sparse symmetric positive definite linear systems and can be easily solved numerically. In the following, we denote the above method for reconstructing a quasi-conformal map $f$ from a Beltrami coefficient $\mu$ by $f = \textbf{LBS}(\mu)$.

\section{Proposed methods}\label{sec:main}
In this section, we first propose a novel algorithm for computing ellipsoidal density-equalizing maps (abbreviated as EDEM) of genus-0 closed surfaces onto a prescribed ellipsoid. Next, we propose another algorithm for computing ellipsoidal density-equalizing quasi-conformal maps (abbreviated as EDEQ), which allows us to simultaneously optimize the shape of the ellipsoidal domain and the mapping onto it to achieve minimal geometric distortions.

\subsection{Ellipsoidal density-equalizing map (EDEM)}\label{sec:EDEM}
Consider a given genus-0 closed surface $\mathcal{M}$ in $\mathbb R^3$ discretized as a triangle mesh $\left(\mathcal{V}, \mathcal{E}, \mathcal{F} \right)$, where $\mathcal{V}$ is the set of vertices, $\mathcal{E}$ is the set of edges which indicate the connections between vertices and are in the form of Euclidean straight line segments in $\mathbb{R}^3$, and $\mathcal{F}$ is the set of triangle faces. Our goal is to compute a bijective ellipsoidal density-equalizing map $f:\mathcal{M} \rightarrow \mathcal{E}_{a,b,c}$, where $\mathcal{E}_{a,b,c}$ is a prescribed ellipsoid with elliptic radii $a,b,c$:
\begin{equation}\label{eqt:ellipsoid}
    \mathcal{E}_{a,b,c} = \left\{(x,y,z) \in \mathbb{R}^3 : \  \frac{x^2}{a^2} + \frac{y^2}{b^2} + \frac{z^2}{c^2} = 1 \right\}.
\end{equation}
Here, the density $\rho$ involved in the density-equalizing mapping method will be defined by some prescribed positive quantity (denoted as the ``population'') per unit area. By changing the ``population'', we can easily control the shape deformation and achieve different desired parameterization results.

\subsubsection{Initial ellipsoidal parameterization}\label{sec:edem_initial}
First, we compute an initial ellipsoidal conformal parameterization $f_{0}:\mathcal{M} \rightarrow \mathcal{E}_{a,b,c}$.

To accomplish this, we apply the Fast Ellipsoidal Conformal Mapping (FECM) method~\cite{choi2024fast}, which is an efficient algorithm for computing an ellipsoidal conformal parameterization by leveraging quasi-conformal theory. Initially, the input genus-0 closed surface $\mathcal{M}$ is conformally mapped to the unit sphere $\mathbb{S}^2$. Then, the FECM method maps the sphere onto the complex plane using the stereographic projection, followed by a M\"obius transformation step that maps two desired polar points of $\mathcal{M}$ to $0$ and $\infty$, respectively. However, it is important to note that the stereographic projection may lead to an uneven distribution of points. To address this issue, the FECM method incorporates an additional rescaling step to enhance the distribution of points. 

Next, the FECM method looks for an inverse projection that maps the planar domain to the target ellipsoid $\mathcal{E}_{a,b,c}$ conformally. It is worth noting that while one can extend the ordinally stereographic projection and its inverse to the ellipsoidal case, the inverse ellipsoidal stereographic projection is not inherently conformal. Therefore, by utilizing the idea of quasi-conformal composition, the FECM method identifies a suitable additional quasi-conformal map on the plane and composes it with the inverse ellipsoidal stereographic projection to form a conformal map from the plane to the ellipsoid $\mathcal{E}_{a,b,c}$. 

Finally, by combining all of the above-mentioned mappings, the FECM method produces an ellipsoidal conformal parameterization $f_{0}: \mathcal{M} \rightarrow \mathcal{E}_{a,b,c}$. Readers are referred to~\cite{choi2024fast} for more details.

\subsubsection{Density-equalizing map on ellipsoid}\label{sec:dem_ellipsoid}
After obtaining an initial ellipsoidal parameterization, we handle the subsequent shape deformation problem on the ellipsoidal domain $\mathcal{E}_{a,b,c}$. 

We define the initial density $\rho$ as the ``population'' per unit area, where the ``population'' is some prescribed positive quantity defined everywhere on the ellipsoid. In the following, we propose an iterative scheme for computing ellipsoidal density-equalizing maps on $\mathcal{E}_{a,b,c}$, with the bijectivity of the mappings enforced. As the density diffusion process is time-dependent, here we denote the position of a vertex $\mathbf{r}$ on $\mathcal{E}_{a,b,c}$ at time $t$ as $\mathbf{r}(t)$. The initial position of $\mathbf{r}$ is represented as $\mathbf{r}(0)$.

Analogous to the original density-equalizing mapping formulation~\cite{gastner2004diffusion}, the diffusion equation of $\rho$ on the ellipsoid is given by:
\begin{equation}\label{eq:diffusion}
    \frac{\partial \rho(\mathbf{r}(t),t)}{\partial t} = \Delta \rho(\mathbf{r}(t),t),
\end{equation}
where $\Delta$ is the Laplace--Beltrami operator. The velocity field induced by the density gradient on the ellipsoid is given by
\begin{equation}\label{eq:velocity}
    \mathbf{v}(\mathbf{r}(t),t) = -\frac{\nabla \rho(\mathbf{r}(t),t)}{\rho(\mathbf{r}(t),t)}.
\end{equation}
Then, the position of any vertex on the ellipsoid can be computed using the following equation:
\begin{equation}\label{eq:displacement}
    \mathbf{r}(t) = \mathbf{r}(0) + \int^{t}_{0}\mathbf{v}(\mathbf{r}(\tau),\tau) d\tau.
\end{equation}
As time progresses towards infinity ($t \rightarrow \infty$), the density is equalized and we obtain the desired ellipsoidal density-equalizing map.

Note that in the discrete case, the initial density $\rho$ needs to be discretized on the ellipsoidal triangle mesh. Here, we first define the initial density $\rho^0_{\mathcal{F}}(T) = \frac{\text{Population}}{\text{Area}(T)}$ on every triangular face $T$ on the ellipsoid. The subsequent iterative process involves updating the positions of all vertices $\left\{ \mathrm{r}_{i}(t_n) \right\}_{i}$, where $i = 1,2,\dots, |\mathcal{V}|$ and at different time points $t_n$, where $n= 1, 2, \dots$ with the time step size $\delta t = t_{n+1}-t_n$. In addition to the face density $\rho^n_{\mathcal{F}}$, we also denote the vertex density as $\rho^n_{\mathcal{V}}$. Note that $\rho^n_{\mathcal{V}}$ can be obtained from $\rho^n_{\mathcal{F}}$ by a simple formula~\cite{choi2018density}:
\begin{equation}\label{eq:conversion}
    \rho^n_{\mathcal{V}} = M\rho^n_{\mathcal{F}},
\end{equation}
where $M$ is a $|\mathcal{V}| \times |\mathcal{F}|$ face-to-vertex conversion matrix such that:
\begin{equation}\displaystyle
    M_{ij} =  \left\{\begin{array}{ll}
\frac{\text{Area}(T_j)}{\sum_{T \in N^{\mathcal{F}}(\mathbf{r}_i)}\text{Area}(T)} & \text{ if $T_j$ is incident to $\mathbf{r}_i$},\\
0 & \text{ otherwise.}
\end{array}\right.
\end{equation}
Here, $N^{\mathcal{F}}(\mathbf{r}_i)$ is the 1-ring neighborhood of the vertex $\mathbf{r}_i$. 

Now, to solve Eq.~\eqref{eq:diffusion} in the discrete case, note that the Laplace-Beltrami operator $\Delta_n$ at the $n$-th iteration can be discretized as:
\begin{equation}
    \Delta_n = -A^{-1}_{n}L_{n}.
\end{equation}
Here, $A_n$ is a diagonal matrix (known as the lumped mass matrix) of size $|\mathcal{V}| \times |\mathcal{V}|$ given by:
\begin{equation}\label{eq:A_n}
    (A_n)_{ii} = \frac{1}{3}\sum_{[\mathbf{r}_i,\mathbf{r}_j,\mathbf{r}_k]\in N^{\mathcal{F}}(\mathbf{r}_i)}\text{Area}\left( [\mathbf{r}_i(t_n),\mathbf{r}_j(t_n),\mathbf{r}_k(t_n)] \right),
\end{equation}
where $[\mathbf{r}_i,\mathbf{r}_j,\mathbf{r}_k]$ is a triangle in $N^{\mathcal{F}}(\mathbf{r}_i)$. It is worth noting that every element of the lumped mass matrix $A_n$ is the area sum of the 1-ring neighborhood of the corresponding vertex at $t = t_n$. The matrix $L_n$ is a $|\mathcal{V}| \times |\mathcal{V}|$ symmetric sparse matrix with cotangent weights~\cite{pinkall1993computing}:
\begin{equation}\label{eq:L_n}
\displaystyle
(L_{n})_{ij} = \left\{\begin{array}{ll}
- \frac{1}{2}(\cot \alpha_{ij} + \cot \beta_{ij}) & \text{ if } [\mathbf{r}_i,\mathbf{r}_j] \in \mathcal{E},\\
- \sum_{k \neq i} (L_n)_{ik} & \text{ if } j = i,\\
0 & \text{ otherwise,}
\end{array}\right.
\end{equation}
where $\alpha_{ij}$ and $\beta_{ij}$ are the two angles opposite to the edge $[\mathbf{r}_i,\mathbf{r}_j]$. Now, Eq.~\eqref{eq:diffusion} can be solved using the semi-discrete backward Euler method:
\begin{equation}
    \frac{\rho^{n+1}_{\mathcal{V}}-\rho^{n}_{\mathcal{V}}}{\delta t} = \Delta_n \rho^{n+1}_{\mathcal{V}},
\end{equation}
from which we have
\begin{equation}\label{eq:diffusion_discrete}
    \rho^{n+1}_{\mathcal{V}} = (A_{n} + \delta t L_{n})^{-1} (A_n \rho^{n}_{\mathcal{V}}).
\end{equation}
After solving Eq.~\eqref{eq:diffusion}, the velocity field can be updated using Eq.~\eqref{eq:velocity}. More specifically, we first compute the discretized density gradient on the triangle mesh. For any triangle element $T = [\mathbf{r}_i,\mathbf{r}_j,\mathbf{r}_k] \in \mathcal{F}$, using the Whitney 0-forms, $\rho^{n+1}_\mathcal{V}$ can be interpolated at any point $x$ on the triangle $T$ by:
\begin{equation}
    \rho^{n+1}_\mathcal{V}(x) = \rho^{n+1}_\mathcal{V}(\mathbf{r}_i)\Psi^{W}_i(x) + \rho^{n+1}_\mathcal{V}(\mathbf{r}_j)\Psi^{W}_j(x) + \rho^{n+1}_\mathcal{V}(\mathbf{r}_k)\Psi^{W}_k(x),
\end{equation}
where $\Psi^{W}_i(x)$, $\Psi^{W}_j(x)$, $\Psi^{W}_k(x)$ are hat functions and $\rho^{n+1}_\mathcal{V}(\mathbf{r}_i)$, $\rho^{n+1}_\mathcal{V}(\mathbf{r}_j)$, $\rho^{n+1}_\mathcal{V}(\mathbf{r}_k)$ are the vertex densities at $\mathbf{r}_i, \mathbf{r}_j, \mathbf{r}_k$. By the property $\nabla \Psi^{W}_i = \frac{N\times e_{ij}}{2\text{Area}(T)}$ (see Ref.~\cite{choi2018density}), where $N$ is the unit normal vector of $T$ and $e_{ij}$ is the directed edge $[\mathbf{r}_i,\mathbf{r}_j]$, the density gradient $\nabla \rho^{n+1}_{\mathcal{F}}(T)$ on the triangle $T$ is given by: 
\begin{equation}
    \nabla \rho^{n+1}_{\mathcal{F}}(T) = \frac{\mathbf{n}_{\mathcal{F}}(T) \times \left(\rho^{n+1}_{\mathcal{V}}(\mathbf{r}_i) e_{jk} + \rho^{n+1}_{\mathcal{V}}(\mathbf{r}_j) e_{ki} + \rho^{n+1}_{\mathcal{V}}(\mathbf{r}_k) e_{ij}\right)}{2\text{Area}(T)},
\end{equation}
where $e_{ij}$, $e_{jk}$, $e_{ki}$ are the three directed edges $[\mathbf{r}_i(t_n),\mathbf{r}_j(t_n)]$, $[\mathbf{r}_j(t_n), \mathbf{r}_k(t_n)]$, $[\mathbf{r}_k(t_n), \mathbf{r}_i(t_n)]$, and $\mathbf{n}_{\mathcal{F}}(T)$ is the outward unit normal vector of the triangle face $T$. Then, using Eq.~\eqref{eq:conversion}, we can obtain $\nabla \rho^{n+1}_{\mathcal{V}}$ on every vertex by:
\begin{equation}\label{eq:gradrho_conversion}
    \nabla \rho^n_{\mathcal{V}} = M \nabla \rho^n_{\mathcal{F}}.
\end{equation}
After obtaining the vertex density $\rho^n_{\mathcal{V}}$ and its gradient $\nabla \rho^n_{\mathcal{V}}$, the velocity at each vertex can be computed using Eq.~\eqref{eq:velocity} as follows:
\begin{equation}\label{eqt:vertex_velocity}
    \mathbf{v}^{n+1}_{\mathcal{V}}(\mathbf{r}_i) = - \frac{\nabla \rho^{n+1}_{\mathcal{V}}(\mathbf{r}_i)}{\rho^{n+1}_{\mathcal{V}}(\mathbf{r}_i)}. 
\end{equation}
However, it is noteworthy that in the discrete case, the velocity in Eq.~\eqref{eqt:vertex_velocity} may not be located in the tangent space. To resolve this issue, we add a projection step to project the velocity onto the admissible space. Since the equation of the ellipsoid $\mathcal{E}_{a,b,c}$ is 
\begin{equation}
    \frac{x^2}{a^2} + \frac{y^2}{b^2} + \frac{z^2}{c^2} = 1,
\end{equation}
it is easy to see that the outward unit normal vector of $\mathcal{E}_{a,b,c}$ is
\begin{equation}
    \mathbf{n}_{\mathcal{V}}\left(\mathbf{r}(t)\right) = \frac{\left(\frac{2x}{a^2},\frac{2y}{b^2},\frac{2z}{c^2} \right)}{\left( \left( \frac{2x}{a^2} \right)^2 + \left( \frac{2y}{b^2} \right)^2 + \left( \frac{2z}{c^2} \right)^2 \right)^{\frac{1}{2}}} = \frac{\left(\frac{x}{a^2},\frac{y}{b^2},\frac{z}{c^2} \right)}{\left( \left( \frac{x}{a^2} \right)^2 + \left( \frac{y}{b^2} \right)^2 + \left( \frac{z}{c^2} \right)^2 \right)^{\frac{1}{2}}}.
\end{equation}
Therefore, the projected velocity field $\widetilde{\mathrm{v}}^{n+1}_{\mathcal{V}}$ can be obtained by
\begin{equation}\label{eqt:projected_velocity}
    \widetilde{\mathbf{v}}^{n+1}_{\mathcal{V}}(\mathbf{r}_i) = \mathbf{v}^{n+1}_{\mathcal{V}}(\mathbf{r}_i) - \left( \mathbf{v}^{n+1}_{\mathcal{V}}(\mathbf{r}_i) \cdot \mathbf{n}_{\mathcal{V}}(\mathbf{r}_i) \right) \mathbf{n}_{\mathcal{V}}(\mathbf{r}_i),
\end{equation}
where $\mathbf{n}_{\mathcal{V}}(\mathbf{r}_i)$ is the outward unit normal vector at the vertex $\mathbf{r}_i$. We can then update the vertex positions based on Eq.~\eqref{eq:displacement}:
\begin{equation}\label{eq:displacement_discrete}
    \mathbf{r}_i(t_{n+1}) = \mathbf{r}_i(t_{n}) - \delta t \ \widetilde{\mathbf{v}}^{n+1}_{\mathcal{V}}(\mathbf{r}_i).
\end{equation}
Note that there may still be small numerical errors that cause the vertices to move outside the ellipsoidal surface. To ensure that all vertices remain on the ellipsoid exactly, we further divide the coordinates of each vertex $\mathbf{r}_i(t_{n+1})$ by $\sqrt{\frac{x^2}{a^2} + \frac{y^2}{b^2} + \frac{z^2}{c^2}}$, where $\mathbf{r}_i(t_{n+1}) = (x,y,z)$. This ensures that all updated vertices satisfy the parametric equation in Eq.~\eqref{eqt:ellipsoid} and hence lie on the ellipsoid $\mathcal{E}_{a,b,c}$.

\subsubsection{Ensuring the bijectivity of the ellipsoidal mapping throughout the iterative process}\label{sec:Ensure_bijectiviy}

As mentioned in~\cite{lyu2024bijective,lyu2024spherical}, conventional density-equalizing mapping methods do not provide any guarantee of bijectivity throughout the diffusion process. Specifically, mesh overlaps may occur under the mapping update in Eq.~\eqref{eq:displacement_discrete} if the density gradient or the time step size is too large. To resolve this issue, we follow the idea in~\cite{lyu2024spherical} and propose a correction scheme that ensures the bijectivity of the ellipsoidal mapping at each iteration.

We first introduce a rescaling transformation $h: \mathcal{E}_{a,b,c} \to \mathbb{S}^2$ defined by: 
\begin{equation}
    h\left(x,y,z \right) = \left(\frac{x}{a}, \frac{y}{b}, \frac{z}{c} \right),
\end{equation}
where $a$, $b$, and $c$ are the radii of the ellipsoid. Using the rescaling transformation $h$, the initial ellipsoidal map $\mathbf{r}(0)$ and the ellipsoidal map $\mathbf{r}(t_n)$ after the $n$-th iterative update can be mapped onto two unit spheres $\mathcal{S}^0 := h(\mathbf{r}(0))$ and $\mathcal{S}^n :=h(\mathbf{r}(t_n))$, respectively. Note that in our work, we consider surfaces discretized in the form of triangle meshes, where the connections between vertices are represented by Euclidean straight line segments in $\mathbb{R}^3$. Therefore, after applying the transformation $h$ on all vertices, we can examine the bijectivity of the mapping with respect to the mesh formed by the vertices and the Euclidean straight line segments. Since $h$ is a piecewise linear anisotropic rescaling map defined on the entire triangle mesh, it is easy to see that $h$ is a bijection and hence will not have any effect on the presence or absence of local mesh fold-overs in the ellipsoids.

Now, we can apply the spherical overlap correction scheme proposed by the SDEM work~\cite{lyu2024spherical}, which is designed for enforcing the bijectivity of a spherical mapping. Specifically, the scheme involves projecting the two spheres $\mathcal{S}^0$ and $\mathcal{S}^n$ onto the extended complex plane via the north-pole stereographic projection and the south-pole stereographic projection and correcting the local mesh fold-overs on the plane using the LBS method~\cite{lui2013texture}. Readers are referred to~\cite{lyu2024spherical} for the detailed description of the spherical overlap correction scheme. We denote the process of the spherical overlap correction scheme as a map $\Tilde{g}_n: \mathcal{S}^n \to \mathbb{S}^2$, where the mapping result $\Tilde{g}_n(\mathcal{S}^n)$ is folding-free.

Finally, we apply the inverse rescaling transformation $h^{-1}$ to rescale the spherical mapping result to the ellipsoid. Again, it is easy to see that $h^{-1}$ is a bijection and will not have any effect on the presence or absence of mesh fold-overs. Therefore, the composition $ h^{-1}\circ \Tilde{g}_n \circ h$ is a mapping from $\mathcal{E}_{a,b,c}$ to $\mathcal{E}_{a,b,c}$ that can correct any local mesh fold-overs in $\mathbf{r}(t_n)$. This completes the ellipsoidal overlap correction scheme.

\subsubsection{Re-coupling the deformation and density}\label{sec:re-couple}
Recall that the proposed ellipsoidal density-equalizing mapping scheme creates shape deformations on a prescribed ellipsoid based on density diffusion. Specifically, the velocity field at each iteration is determined by the density and its gradient at the mesh vertices. As the computational procedure involves multiple discretization schemes, numerical errors may accumulate throughout the iterations. Also, the above-mentioned overlap correction scheme may alter the vertex positions when correcting the local mesh fold-overs, thereby leading to a discrepancy between the vertex positions and the actual density flow. To address this issue, we follow the idea in~\cite{lyu2024spherical} and introduce an extra step to re-couple the density and deformation at the end of every iteration. 

More specifically, at the end of the $n$-th iteration, we do not directly use the density obtained by solving the diffusion equation in Eq.~\eqref{eq:diffusion_discrete} to continue the next iteration. Instead, we re-define the density $\rho^{n+1}_{\mathcal{F}}(T)$ on the triangle element $T = [\mathbf{r}_i,\mathbf{r}_j,\mathbf{r}_k]$ using the updated vertex positions $\mathbf{r}(t_{n+1})$ on the ellipsoid:
\begin{equation}\label{eq:re-couple}
    \rho^{n+1}_{\mathcal{F}}(T) = \frac{\text{Population}}{\text{Area}[\mathbf{r}_i(t_{n+1}),\mathbf{r}_j(t_{n+1}),\mathbf{r}_k(t_{n+1})]}.
\end{equation}
Once the updated density $\rho^{n+1}_{\mathcal{F}}$ is obtained, we calculate $\rho^{n+1}_{\mathcal{V}}$ using Eq.~\eqref{eq:conversion} and proceed with the next iteration. Altogether, this additional re-coupling step allows us to reduce the accumulation of numerical errors as well as the discrepancy between the shape deformation and the density.

\subsubsection{Summary}
By integrating the initial ellipsoidal conformal parameterization (Section~\ref{sec:edem_initial}) and the diffusion iteration scheme (Section~\ref{sec:dem_ellipsoid}) with the overlap correction scheme (Section~\ref{sec:Ensure_bijectiviy}) and the re-coupling scheme (Section~\ref{sec:re-couple}), we have the proposed EDEM algorithm for computing bijective ellipsoidal density-equalizing maps for genus-0 closed surfaces. The proposed algorithm is summarized in Algorithm~\ref{alg:EDEM}. In practice, we set the step size $\delta t = 0.1$, the stopping parameter $\epsilon = 10^{-3}$, and the maximum number of iterations allowed $n_{\max} = 300$. 

\begin{algorithm}[h!]
\KwIn{A genus-0 closed surface $\mathcal{M}$, a prescribed population, the elliptic radii $a,b,c$, the step size $\delta t$, the stopping parameter $\epsilon$, and the maximum number of iterations allowed $n_{\max}$.}
\KwOut{An ellipsoidal density-equalizing map $f:\mathcal{M}\to \mathcal{E}_{a,b,c}$.}
\BlankLine

Compute an initial ellipsoidal conformal parameterization $f_0:\mathcal{M} \to \mathcal{E}_{a,b,c}$~\cite{choi2024fast}\;

Compute the initial density $\rho^0_{\mathcal{F}}$ on $f_0(\mathcal{M})$ based on the prescribed population\;

Set $n = 0$\;

\Repeat{$\frac{\text{sd}\left(\rho^{n}_{\mathcal{V}}\right)}{\text{mean}\left(\rho^{n}_{\mathcal{V}}\right)} < \epsilon$ \ or \ $n \geq n_{\max}$}{ 

Compute $A_n$ and $L_n$ on $\mathcal{E}_{a,b,c}$ by Eq.~\eqref{eq:A_n} and Eq.~\eqref{eq:L_n}\;

Obtain $\rho^{n+1}_{\mathcal{V}}$ by solving the diffusion equation Eq.~\eqref{eq:diffusion}\;

Compute the velocity field ${\mathbf{v}}^{n+1}_{\mathcal{V}}$ at every vertex by Eq.~\eqref{eqt:vertex_velocity}\;

Compute the projected velocity field $\widetilde{\mathbf{v}}^{n+1}_{\mathcal{V}}$ onto the ellipsoid using Eq.~\eqref{eqt:projected_velocity}\;

Update the position of all vertices using Eq.~\eqref{eq:displacement_discrete} and enforce the vertices to remain on $\mathcal{E}_{a,b,c}$\;

Apply the overlap correction scheme in Section~\ref{sec:Ensure_bijectiviy}\;

Apply the re-coupling scheme in Section~\ref{sec:re-couple} to update $\rho^{n+1}_{\mathcal{F}}$\;

Update $n = n + 1$\;
}

The resulting bijective ellipsoidal density-equalizing map is $f: \mathcal{M} \rightarrow \mathcal{E}_{a,b,c} $ \;
\caption{Ellipsoidal density-equalizing map (EDEM)}
\label{alg:EDEM}
\end{algorithm}

\subsection{Ellipsoidal density-equalizing quasi-conformal map (EDEQ)}\label{sec:optimizeabc}
While our EDEM algorithm can produce shape deformations and achieve prescribed area changes on the given ellipsoid $\mathcal{E}_{a,b,c}$, other geometric distortions such as conformal distortions may be uncontrolled. Also, note that the EDEM method can be applied to ellipsoids with any prescribed elliptic radii. It is therefore natural to ask whether one can further reduce the overall geometric distortions by optimizing both the shape of the ellipsoid and the associated ellipsoidal mappings. 

As described in the DEQ method~\cite{lyu2024bijective}, the density-equalizing process can be considered as an energy minimization problem involving the prescribed density. Analogously, as our EDEM method aims to create a shape deformation following the density diffusion process on the ellipsoid, the density-equalizing effect of a map $f$ can be assessed using the following energy:
\begin{equation}
    E_{\text{EDEM}}(f) = \int \left\| \frac{\nabla \rho}{\rho}\right\|^2,
\end{equation}
where $\rho$ is the density associated with the map $f$. Also, note that by quasi-conformal theory, the conformal distortion of a map $f$ can be represented using the norm of its Beltrami coefficient $\|\mu\|$, where a smaller $\|\mu\|$ implies a smaller eccentricity of the local ellipses and hence a smaller angle distortion. Therefore, we can assess the conformal distortion of a mapping by considering the following energy:
\begin{equation}
    E_{\text{BC}}(f) = \int |\mu |^2. 
\end{equation}

In this section, our goal is to develop an ellipsoidal density-equalizing quasi-conformal mapping algorithm, which we abbreviate as EDEQ, to reduce both of the above energies $E_{\text{EDEM}}$ and $E_{\text{BC}}$. Now, it is noteworthy that the shape of the ellipsoid $\mathcal{E}_{a,b,c}$, characterized by the elliptic radii $a,b,c$, can effectively provide us with extra degrees of freedom in reducing the energies. Therefore, we consider minimizing the following combined energy $E$ with the radii $a,b,c$ and the map $f: \mathcal{M} \to \mathcal{E}_{a,b,c}$ being the variables:
\begin{equation}\label{eq:combined_energy}
    E(a,b,c,f)  = E_{\text{EDEM}} + \alpha E_{\text{BC}} = \int \left\|\frac{\nabla \rho}{\rho} \right\|^2 + \alpha \int |\mu |^2,
\end{equation}
where $\alpha$ is a nonnegative weighting parameter for balancing the density-equalizing error and the conformal distortion. The existence of the minimizer of $E$ can be proved by following the arguments in~\cite{lyu2024bijective,lui2007landmark}.

\subsubsection{Decoupling the combined energy}
To simplify the optimization process, we first decouple the minimization problem of the combined energy $E$ into two subproblems of minimizing $E_1$ and $E_2$ below:
\begin{equation}\label{eq:density_part}
    E_1(f) = \int_{\mathcal{E}_{a,b,c}} \left\|\frac{\nabla \rho}{\rho} \right\|^2,  
\end{equation}
\begin{equation}\label{eq:optimize_shape}
    E_2(a,b,c) = \int_{\mathcal{E}_{a,b,c}}  \left\|\frac{\nabla \rho}{\rho} \right\|^2 + \alpha \int_{\mathcal{E}_{a,b,c}} |\mu|^2.
\end{equation}
More specifically, the energy $E_1$ in Eq.~\eqref{eq:density_part} aims to equalize the density distribution across a fixed ellipsoidal domain with radii $a,b,c$, promoting a more uniform density throughout the iterations. The energy $E_2$ in Eq.~\eqref{eq:optimize_shape} focuses on optimizing the shape of the ellipsoid to minimize the combined geometric distortion. We can then utilize our developed computational scheme in the EDEM method for the subproblem $E_1$ and focus on updating the ellipsoidal geometry in the subproblem $E_2$.

\subsubsection{Subproblem $E_1$}\label{sec:descent_direction_E1}
We first consider the descent direction of $E_1 = \int_{\mathcal{E}_{a,b,c}} \left\|\frac{\nabla \rho}{\rho} \right\|^2$, with the elliptic radii $a,b,c$ fixed. As in the formulation of the EDEM method in Section~\ref{sec:EDEM}, the velocity field can be computed by $\mathbf{v} = -\frac{ \nabla \rho}{\rho}$. Also, to preserve the given ellipsoidal shape, we remove the normal component $\mathbf{v}^{\perp} = \left(\mathbf{v}\cdot \mathbf{n} \right)\mathbf{n}$, where $\mathbf{n}$ is the outward normal unit vector. The descent direction of $E_1$ is then given by:
\begin{equation}\label{eq:dE_1}
    dE_1 = \mathbf{v} - \mathbf{v}^{\perp}.
\end{equation}
Hence, we have the following iterative scheme for the energy $E_1$: 
\begin{equation}
    f_{n+1} = f_n + \delta t dE^n_1,
\end{equation}
where $\delta t$ is the time step size, $f_n$ is the map at the $n$-th iteration, and $dE^n_1$ is the descent direction of $E_1$ at the $n$-th iteration. 

\subsubsection{Subproblem $E_2$}\label{sec:shape_optimization}
To minimize the energy $E_2$, we first consider the Beltrami coefficient term $\int_{\mathcal{E}_{a,b,c}} |\mu|^2$. Recall that the Beltrami coefficient is a complex-valued function defined on the complex plane. To extend it for assessing the conformal distortion of an ellipsoidal map, we use a triangle-based approach as follows. Let $T^0= [\mathbf{r}_i(t_0), \mathbf{r}_j(t_0), \mathbf{r}_k(t_0)]$ be a triangle element on the initial ellipsoidal conformal parameterization and $T^n = [\mathbf{r}_i(t_n), \mathbf{r}_j(t_n), \mathbf{r}_k(t_n)]$ be the corresponding triangle at the $n$-th iteration. We rigidly map both $T^0$ and $T^n$ onto the complex plane. Specifically, this can be done by fixing one of the vertices of each triangle at the origin, one of the triangle edges containing this vertex onto the real axis, and then computing the position of the remaining vertex based on the edge lengths and angles of the triangle. We denote the resulting triangles on the complex plane as $\widetilde{T}^0$ and $\widetilde{T}^n$. Then, we can consider the mapping between $\widetilde{T}^0$ and $\widetilde{T}^n$ as a quasi-conformal map $f_{\widetilde{T}^n} = u + iv$. The Beltrami coefficient of $f_{\widetilde{T}^n}$ can be obtained by:
\begin{equation}
    \mu_{\widetilde{T}^n} = \frac{\left(u_x - v_y \right) + i\left(v_x + u_y \right)}{\left(u_x + v_y \right) + i\left(v_x - u_y \right)}.
\end{equation}
From the above formula, we can get the norm of the Beltrami coefficient $\|\mu_{\widetilde{T}^n}\|$ and use it to represent the conformal distortion between $T^0$ and $T^n$. We can then repeat the above procedure for all triangular faces of the surface to get the term $\int_{\mathcal{E}_{a,b,c}} |\mu |^2$.

Next, we consider optimizing the shape of the ellipsoid (i.e. the radii $a,b,c$) based on the energy $E_2$. To simplify the formulation, we keep $a$ unchanged and focus on optimizing the two other elliptic radii $b$ and $c$ using an iterative procedure. We first define the radius step sizes along $b$ and $c$ as $\delta b$ and $\delta c$, respectively. Now, there are nine possible combinations of the radii in the form of $(a, b + k_b \delta b, c + k_c \delta c)$ with $k_b, k_c = 0, 1, -1$. We can then consider the following energies associated with the nine combinations, including the original one:
\begin{equation}
    \begin{array}{l}
        E_2(a,b,c) = \displaystyle \int_{\mathcal{E}_{a,b,c}}  \left( \left\|\frac{\nabla \rho}{\rho}(a,b,c) \right\|^2 + \alpha \left| \mu(a,b,c) \right|^2 \right),
    \end{array}
\end{equation}
the ones with only one among $b$ and $c$ changed:
\begin{equation}
    \begin{array}{l}
        \displaystyle E_2(a,b+\delta b,c) = \int_{\mathcal{E}_{a,b + \delta b,c}} \left( \left\|\frac{\nabla \rho}{\rho}(a,b+\delta b,c) \right\|^2 + \alpha | \mu(a,b+\delta b,c)|^2\right),  \\
        \displaystyle E_2(a,b,c + \delta c) = \int_{\mathcal{E}_{a,b,c + \delta c}} \left( \left\|\frac{\nabla \rho}{\rho}(a,b,c + \delta c) \right\|^2 + \alpha | \mu(a,b,c + \delta c)|^2\right), \\
        \displaystyle E_2(a,b - \delta b,c) = \int_{\mathcal{E}_{a,b - \delta b,c}}  \left(\left\|\frac{\nabla \rho}{\rho}(a,b - \delta b,c) \right\|^2 + \alpha | \mu(a,b - \delta b,c) |^2\right), \\
        \displaystyle E_2(a,b,c - \delta c) = \int_{\mathcal{E}_{a,b,c - \delta b}} \left( \left\|\frac{\nabla \rho}{\rho}(a,b,c - \delta c) \right\|^2 + \alpha | \mu(a,b,c - \delta c)|^2\right), 
    \end{array}
\end{equation}  
and the ones with both $b$ and $c$ changed:
\begin{equation}
    \begin{array}{l}
        \displaystyle E_2(a,b + \delta b,c + \delta c) = \int_{\mathcal{E}_{a,b + \delta b,c + \delta c}}  \left(\left\|\frac{\nabla \rho}{\rho}(a,b + \delta b,c + \delta c) \right\|^2 + \alpha | \mu(a,b + \delta b,c + \delta c) |^2\right),\\
        \displaystyle E_2(a,b + \delta b,c - \delta c) = \int_{\mathcal{E}_{a,b + \delta b,c - \delta c}}  \left(\left\|\frac{\nabla \rho}{\rho}(a,b + \delta b,c - \delta c) \right\|^2 + \alpha | \mu(a,b + \delta b,c - \delta c) |^2\right), \\
        \displaystyle E_2(a,b - \delta b,c + \delta c) = \int_{\mathcal{E}_{a,b - \delta b,c + \delta c}}  \left(\left\|\frac{\nabla \rho}{\rho}(a,b - \delta b,c + \delta c) \right\|^2 + \alpha | \mu(a,b - \delta b,c + \delta c) |^2\right), \\
        \displaystyle E_2(a,b - \delta b,c - \delta c) = \int_{\mathcal{E}_{a,b - \delta b,c - \delta c}}  \left(\left\|\frac{\nabla \rho}{\rho}(a,b - \delta b,c - \delta c) \right\|^2 + \alpha | \mu(a,b - \delta b,c - \delta c)|^2\right).
    \end{array}
\end{equation}
Here, the terms $\frac{\nabla \rho}{\rho}(\Tilde{a},\Tilde{b},\Tilde{c})$ and $\mu(\Tilde{a},\Tilde{b},\Tilde{c})$ are computed based on the updated ellipsoid $\mathcal{E}_{\Tilde{a},\Tilde{b},\Tilde{c}}$. By selecting the combination with the smallest energy value, we can update the shape of the ellipsoid and reduce $E_2$. The shape update procedure is summarized in Algorithm~\ref{alg:optimize_shape}.

\begin{algorithm}[h!]
\KwIn{A ellipsoid $\mathcal{E}_{a,b,c}$, a prescribed population, and the step sizes $\delta b$ and $\delta c$.}
\KwOut{An ellipsoidal triangle mesh $\mathcal{E}_{\Tilde{a},\Tilde{b},\Tilde{c}}$.}
\BlankLine

Compute the energies $E_2(a,b,c)$, $E_2(a,b+\delta b,c)$, $E_2(a,b,c + \delta c)$, $E_2(a,b - \delta b,c)$, $E_2(a,b,c - \delta c)$, $E_2(a,b + \delta b,c + \delta c)$, $E_2(a,b + \delta b,c - \delta c)$, $E_2(a,b - \delta b,c + \delta c)$, $E_2(a,b - \delta b,c - \delta c)$\;

Select $(\Tilde{a},\Tilde{b},\Tilde{c})$ that gives the smallest energy\;

Update the ellipsoid as $\mathcal{E}_{\Tilde{a},\Tilde{b},\Tilde{c}}$ \;

\caption{Shape update of the ellipsoidal domain}
\label{alg:optimize_shape}
\end{algorithm}

\subsubsection{Summary}
In Sections \ref{sec:descent_direction_E1} and \ref{sec:shape_optimization}, we discussed how to solve the subproblems $E_1$ and $E_2$. Combining these approaches, we have our proposed ellipsoidal density-equalizing quasi-conformal map (EDEQ) algorithm. 

Specifically, analogous to the EDEM algorithm, here we start by computing an initial ellipsoidal conformal parameterization (Section~\ref{sec:edem_initial}) with radii $(a,b,c)$. The initial density and the initial combined energy can be obtained using the initial ellipsoidal parameterization. Then, to solve the subproblem $E_1$, we adjust the vertex position on the current ellipsoid based on the descent direction (Section~\ref{sec:descent_direction_E1}) iteratively. The overlap correction scheme (Section~\ref{sec:Ensure_bijectiviy}) and the re-coupling scheme (Section~\ref{sec:re-couple}) are also applied at each iteration. We repeat the above process for a certain number of iterations (set by a prescribed parameter $K$) so that the energy $E_1$ defined based on the initial elliptic radii is sufficiently reduced. We then include an additional shape update step (Section~\ref{sec:shape_optimization})  after every $K$ iteration to modify the ellipsoidal radii by solving the subproblem $E_2$. With the elliptic radii updated, we solve the subproblem $E_1$ to adjust the vertex position on the updated ellipsoid as described above again and repeat the above process. Also, note that a finer adjustment of the elliptic radii will be needed as the iterations continue. Therefore, instead of using fixed radius step sizes $\delta b, \delta c$ in the shape update algorithm, we add a scaling factor $0.9^m$ to the prescribed step sizes $\delta b, \delta c$ at the $m$-th shape update step. We repeat the iterations until $|\frac{E^{n+1} - E^n}{E^n}| < \epsilon$ for some stopping parameter $\epsilon$. The final EDEQ map is obtained by $f = f_n$, from which we also get the final optimal elliptic radii $(a,b,c)$.

The proposed EDEQ algorithm is summarized in Algorithm~\ref{alg:EDEQ}. In practice, the initial radii $(a_0,b_0,c_0)$ can be chosen arbitrarily, the time step size is set to be $\delta t = 0.1$, the initial radius step sizes are set to be $\delta b = 0.1$ and $\delta c = 0.1$, the number of iterations for each set of fixed radii is set to be $K = 5$, the error threshold is set to be $\epsilon = 10^{-5}$, and the maximum number of iterations is set to be $n_{\text{max}} = 300$, which is sufficient for ensuring satisfactory convergence in our experiments.

\begin{algorithm}[h]
\KwIn{A genus-0 closed surface $\mathcal{M}$, a prescribed population, the initial elliptic radii $(a_0,b_0,c_0)$, the step size $\delta t$, the radius step sizes $\delta b$ and $\delta c$,  the number of iterations for each set of fixed radii $K$, the stopping parameter $\epsilon$, and the maximum number of iterations allowed $n_{\max}$.}
\KwOut{An ellipsoidal density-equalizing quasi-conformal map $f:\mathcal{M}\to \mathcal{E}_{a,b,c}$ with the optimal ellipsoid $\mathcal{E}_{a,b,c}$.} 
\BlankLine

Set $(a,b,c) = (a_0, b_0, c_0)$\; 

Compute an initial ellipsoidal conformal parameterization $f_0:\mathcal{M} \to \mathcal{E}_{a,b,c}$~\cite{choi2024fast}\;

Compute the initial density $\rho^0_{\mathcal{F}}$ on $f_0(\mathcal{M})$ based on the prescribed population\;

Set $n = 0$\;

\Repeat{$|\frac{E^{n+1} - E^n}{E^n}| < \epsilon$ \ or \ $n \geq n_{\max}$}{ 

Compute the descent direction $dE^n_1$ by Eq.~\eqref{eq:dE_1}\;

Update the mapping $f_{n+1} = f_n + \delta t dE^n_1$\;

Apply the overlap correction scheme in Section~\ref{sec:Ensure_bijectiviy}\;

Apply the re-coupling scheme in Section~\ref{sec:re-couple} to update $\rho^{n+1}_{\mathcal{F}}$\;

\If{$n =km$ for some positive integer $m$}{
Apply the shape update method (Algorithm~\ref{alg:optimize_shape}) with the radius step sizes $0.9^{m}\delta b$ and $0.9^{m}\delta c$ and obtain the updated elliptic radii $(\Tilde{a},\Tilde{b},\Tilde{c})$\;

Set $(a,b,c) = (\Tilde{a},\Tilde{b},\Tilde{c})$\; 

}

Update $n = n + 1$\;
}

The resulting bijective ellipsoidal density-equalizing quasi-conformal map is $f: \mathcal{M} \rightarrow \mathcal{E}_{a,b,c} $, where $(a,b,c)$ are the optimal elliptic radii\;
\caption{Ellipsoidal density-equalizing quasi-conformal map (EDEQ)}
\label{alg:EDEQ}
\end{algorithm}

\section{Experiments}\label{sect:experiment}

In this section, we present experimental results to demonstrate the effectiveness of our proposed EDEM and EDEQ algorithms. The algorithms are implemented using MATLAB R2021a on the Windows platform. All experiments are conducted on a computer with an Intel(R) Core(TM) i9-12900 2.40 GHz processor and 32GB memory. The surface meshes are from online mesh repositories~\cite{common}. All surfaces are discretized in the form of triangular meshes.

\subsection{Ellipsoidal density-equalizing map}
\begin{figure}[t]
    \centering
    \includegraphics[width=\textwidth]{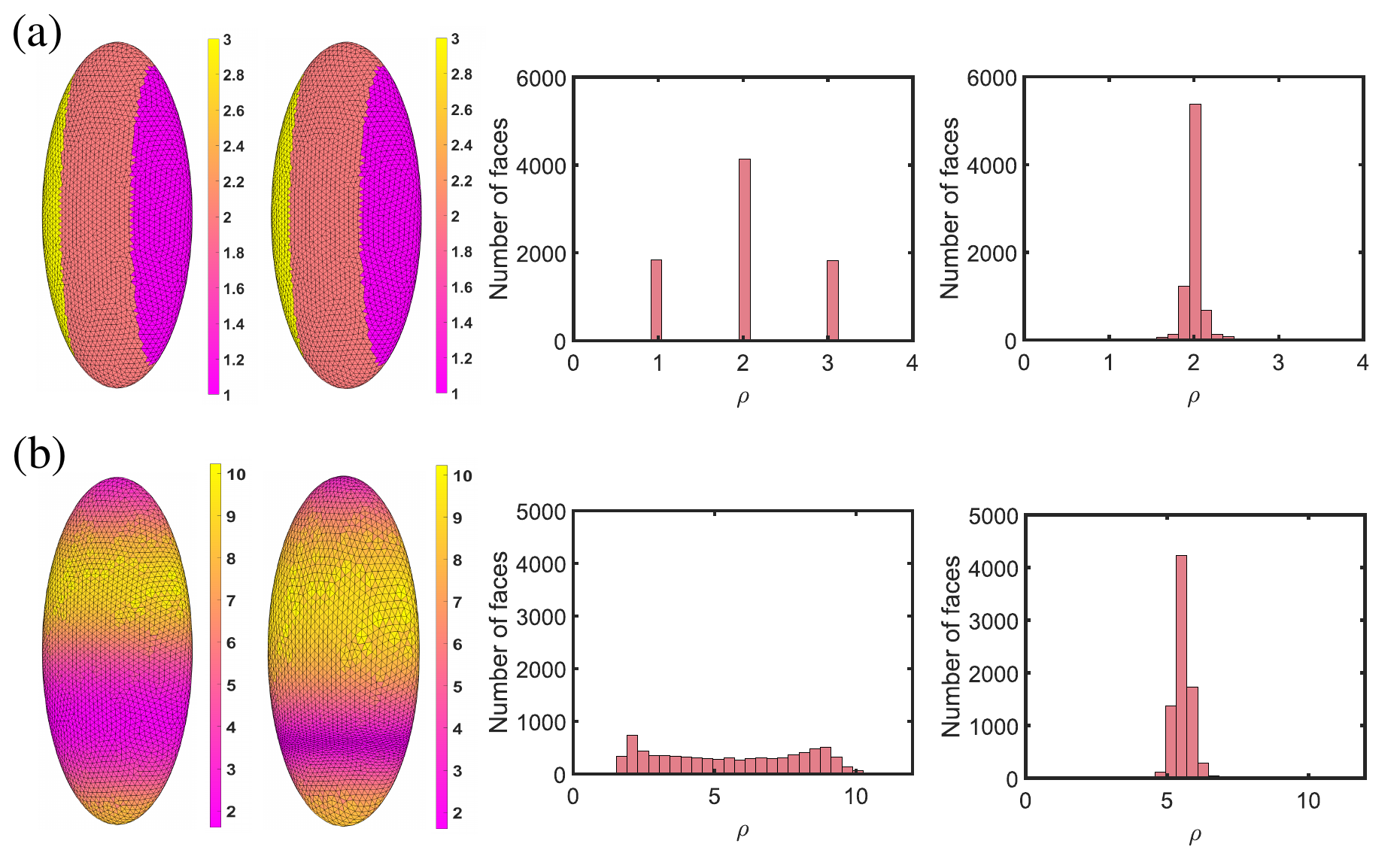}
    \caption{\textbf{Ellipsoidal density-equalizing maps of ellipsoidal surfaces.} Each row shows one example. (a)~An example with discontinuous input density. (b)~An example with continuous input density. Left to right: The initial ellipsoidal surface color-coded with the initial density, the final EDEM result color-coded with the initial density, the histogram of the initial density, and the histogram of the final density. The elliptic radii are $(a,b,c) = (1,2,4)$ for both examples.}
    \label{fig:edem_ellipsoid}
\end{figure}

To test our proposed EDEM algorithm, we first consider mapping an ellipsoidal surface with two different density distributions, including a discontinuous density distribution (Fig.~\ref{fig:edem_ellipsoid}(a)) and a continuous density distribution (Fig.~\ref{fig:edem_ellipsoid}(b)). In both examples, it can be observed that the input densities are highly non-uniform. By applying the proposed EDEM method, we obtain the corresponding ellipsoidal density-equalizing maps. It can be observed that the domains with a high initial density are enlarged while the domains with a low initial density are shrunk, which shows that our method can accurately produce shape deformations on the ellipsoid based on the prescribed density. Also, from the histograms of the initial and final densities, it can be observed that the density is effectively equalized.

\begin{figure}[t!]
    \centering
    \includegraphics[width=\textwidth]{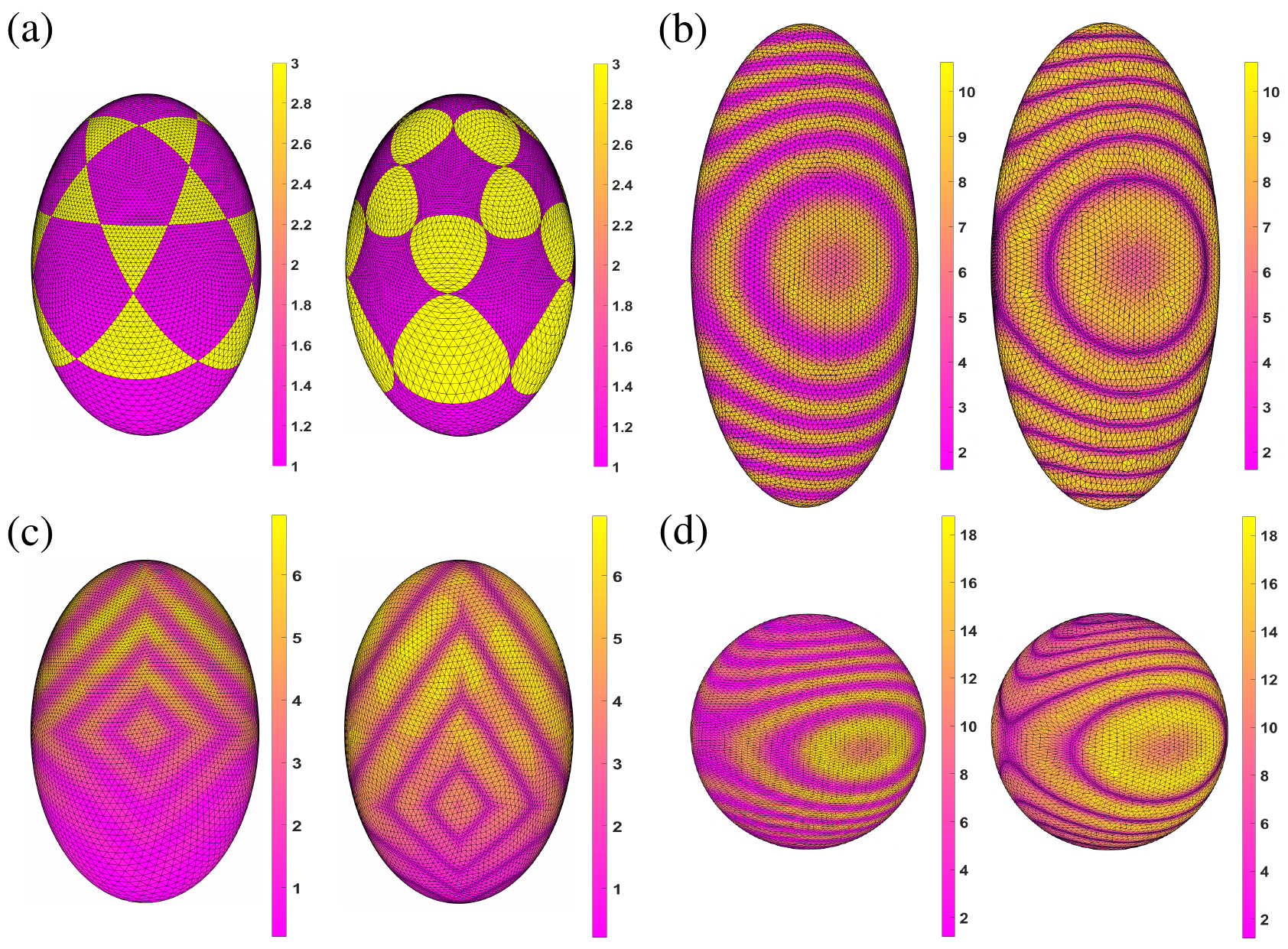}
    \caption{\textbf{Additional examples of ellipsoidal density-equalizing maps of ellipsoidal surfaces.} For each example, the left figure shows the input surface color-coded with the initial density, and the right figure shows the EDEM result color-coded with the initial density. (a)~An ellipsoid with non-uniformly distributed mesh elements and a prescribed discontinuous density. Here, the elliptic radii are $(a,b,c) = (1,1,1.5)$. (b)~An elongated ellipsoid with a prescribed continuous density. Here, the elliptic radii are $(a,b,c) = (1,0.6,2)$. (c)~An ellipsoid with non-uniformly distributed mesh elements and a complex prescribed density. Here the radii are $(a,b,c) = (1,1,1.5)$. (d)~A sphere with a prescribed continuous density. Here, the elliptic radii are $(a,b,c) = (1,1,1)$. }
    \label{fig:sythetic_edem_add}
\end{figure}

Next, we consider some more complicated ellipsoidal mapping examples. Fig.~\ref{fig:sythetic_edem_add}(a) shows an ellipsoidal mesh with non-uniform distributed triangle elements. Specifically, the mesh is denser at the top part of the ellipsoid and much coarser at the bottom part. We define different populations on the ellipsoid such that the initial density of all the triangular regions is three times the density of all the pentagonal regions. From our EDEM mapping result, it can be observed that different regions are enlarged or shrunk accordingly. This shows that our method is capable of computing ellipsoidal density-equalizing maps on a non-uniform mesh. One may also wonder whether our method can be applied to more extreme ellipsoidal geometries and more extreme density distributions. In Fig.~\ref{fig:sythetic_edem_add}(b), we consider a more elongated ellipsoid with a more extreme density distribution prescribed on it, where the maximum and minimum density values differ by 10 times. It can be observed from our EDEM mapping result that different regions are effectively enlarged or shrunk based on the input density. This suggests that our method is capable of handling the extreme ellipsoidal geometry and extreme input density. Fig.~\ref{fig:sythetic_edem_add}(c) shows another ellipsoid with non-uniform mesh elements and a complicated continuous density distribution. Again, we can see that our proposed method produces an ellipsoidal density-equalizing map successfully. Lastly, in Fig.~\ref{fig:sythetic_edem_add}(d) we test our EDEM algorithm on a spherical surface (i.e. an ellipsoid with all three elliptic radii being identical). It can be observed from the mapping result that our method also works well for the spherical case.

For a more quantitative analysis, we record the computational time, radii of the ellipsoid, variance of the initial density, variance of the final density, and number of overlaps for all the above ellipsoidal mapping examples in Table~\ref{tab:EDEM}. It can be observed that our EDEM method can be effectively applied to a wide range of ellipsoidal shapes with different elliptic radii. Specifically, for ellipsoids with different radii and input densities, the variances of the final density in the EDEM result are all close to 0. This shows that all EDEM results are highly density-equalizing. Also, it can be observed that all mapping results are folding-free. Overall, the experiments have demonstrated the versatility of the EDEM method.

\begin{table}[t!]
\small
    \caption{\textbf{The performance of our EDEM algorithm.} For each surface, we record the number of triangle elements, the computational time, the elliptic radii $(a,b,c)$ of the ellipsoidal domain, the variance of the normalized initial density $\widetilde{\rho}_1 = \frac{\rho_1}{\text{Mean}(\rho_1)}$ (where $\rho_1$ is the initial vertex density), the variance of the normalized final density $\widetilde{\rho}_2  = \frac{\rho_2}{\text{Mean}(\rho_2)}$ (where $\rho_2$ is the final vertex density), and the number of overlaps.} \label{tab:EDEM}
    
  \begin{center}
\resizebox{\textwidth}{!}{
  \begin{tabular}{|c|c|c|c|c|c|c|} \hline
    \bf Surface & \bf \# Faces & \bf Time (s) & $\mathbf{(a,b,c)}$ &\bf $\text{Var}(\widetilde{\rho}_1)$  &\bf $\text{Var}(\widetilde{\rho}_2)$ & \bf \# Overlaps \\\hline
    Ellipsoid 1 (Fig.~\ref{fig:edem_ellipsoid}(a)) & 7808 & 1.8171 & (1,2,4) & 0.1176 & 0.0032 & 0 \\ \hline
    Ellipsoid 2 (Fig.~\ref{fig:edem_ellipsoid}(b)) & 7808 & 4.6973 & (1,2,4)  & 0.2084 & 0.0028 & 0 \\ \hline
    Ellipsoid 3 (Fig.~\ref{fig:sythetic_edem_add}(a)) & 20480 & 1.4869
 & (1,1,1.5) & 0.3333 & 0.0137 & 0 \\ \hline
    Ellipsoid 4 (Fig.~\ref{fig:sythetic_edem_add}(b)) & 18904 & 0.3811
 &(1,0.6,2) & 0.2135 & 0.0055 & 0 \\ \hline
    Ellipsoid 5 (Fig.~\ref{fig:sythetic_edem_add}(c)) & 20480 & 2.6060
 & (1,1,1.5) & 0.2580 & 0.0010 & 0  \\ \hline
    Ellipsoid 6 (Fig.~\ref{fig:sythetic_edem_add}(d)) & 18904 & 0.6203
 &(1,1,1) & 0.2940 & 0.0068 & 0 \\ \hline
  \end{tabular}
  }
\end{center}
\end{table}

\begin{figure}[t!]
    \centering
    \includegraphics[width=\textwidth]{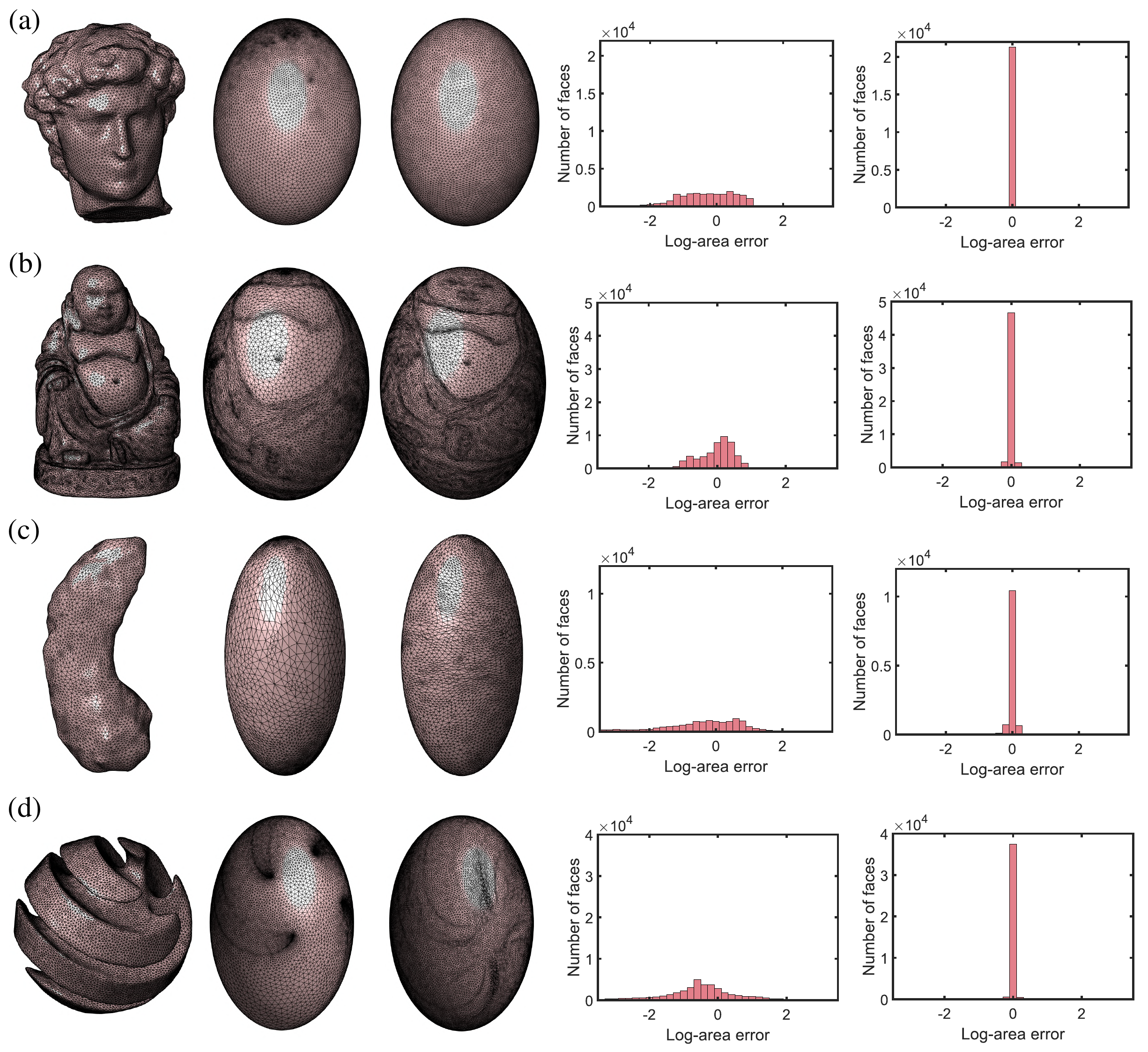}
    \caption{\textbf{Ellipsoidal area-preserving parameterization of genus-0 closed surfaces obtained by the EDEM method.} Each row shows one example. (a)~The David model with elliptic radii $(a,b,c) = (1,1.2,1.4)$. (b)~The Buddha model with elliptic radii $(a,b,c) = (1,1,1.4)$. (c)~The hippocampus model with elliptic radii $(a,b,c) = (1,1,2.2)$. (d)~The twisted ball model with elliptic radii $(a,b,c) = (1,0.8,1.4)$. Left to right: The input surface mesh, the initial ellipsoidal conformal parameterization obtained by the FECM method~\cite{choi2024fast}, the EDEM result, the histogram of the log-area error $d_{\text{area}}$ of the initial ellipsoidal parameterization, and the histogram of the log-area error $d_{\text{area}}$ of the final ellipsoidal parameterization.}
    \label{fig:edem_area}
\end{figure}

Next, we consider computing the ellipsoidal density-equalizing maps for more general genus-0 closed surfaces. In Fig.~\ref{fig:edem_area}, we set the population as the face area of the original mesh and apply our EDEM algorithm to achieve ellipsoidal area-preserving parameterizations. For the initial parameterization step, we map each surface conformally onto the ellipsoid using the FECM method~\cite{choi2024fast}. Then, we apply the EDEM method to compute the ellipsoidal density-equalizing parameterizations. To assess the area-preserving property of the ellipsoidal parameterizations, we consider the log-area error for each triangular face $T$ as follows:
\begin{equation}\label{eq:log_area}
    d_{\text{area}}(f)(T) = \log\left(\frac{\text{Area}(f(T))/\sum_{T'\in \mathcal{F}} \text{Area}(f(T'))}{\text{Area}(T)/\sum_{T'\in \mathcal{F}} \text{Area}(T')}\right).
\end{equation}
Here $f$ denotes the ellipsoidal parameterization and $\mathcal{F}$ is the set of all triangular faces. The normalization factors in the numerator and denominator ensure that a perfectly area-preserving mapping would result in $d_{\text{area}} \equiv 0$. In Fig.~\ref{fig:edem_area}, we present various genus-0 surface models, the initial ellipsoidal parameterizations, the ellipsoidal parameterizations obtained by our EDEM method, and the initial and final log-area errors. It can be observed that the EDEM method performs very well for genus-0 closed surfaces with complex structures. Specifically, by visually comparing the triangle element distributions in the initial ellipsoidal parameterizations and the EDEM parameterization results, it is easy to see that the EDEM method achieves a more uniform distribution on the prescribed ellipsoid. By comparing the initial and final log-area error histograms, we can also see that the log-area error is significantly reduced by our EDEM method. For a more quantitative analysis, Table~\ref{tab:EDEM_area} provides detailed statistics on the performance of our EDEM algorithm for ellipsoidal area-preserving parameterization. We can see that the EDEM method can efficiently reduce the log-area error by over 95\% on average when compared to the initial parameterization, and the bijectivity of the parameterization is well-preserved in all examples. Altogether, the experiments show that our EDEM method is capable of computing bijective ellipsoidal area-preserving parameterizations for a large variety of genus-0 closed surfaces.

\begin{table}[t!]
\small
    \caption{\textbf{The performance of our EDEM algorithm for ellipsoidal area-preserving parameterization.} For each surface, we record the number of triangle elements, the computational time, the elliptic radii $(a,b,c)$ of the target ellipsoidal domain, mean and standard deviation of the initial log-area error $|d_{\text{area}}(f_0)|$ and the final log-area error $|d_{\text{area}}(f)|$, and the number of overlaps.}\label{tab:EDEM_area}
    
  \begin{center}
\resizebox{\textwidth}{!}{
  \begin{tabular}{|C{22mm}|c|c|c|c|c|c|c|c|c|} \hline
    \multirow{ 2}{*}{\bf Surface} & \multirow{ 2}{*}{\bf \# Faces} & \multirow{ 2}{*}{\bf Time (s)} & \multirow{ 2}{*}{$\mathbf{(a,b,c)}$} & \multicolumn{2}{c|}{\bf $|d_{\text{area}}(f_0)|$} & \multicolumn{2}{c|}{\bf $|d_{\text{area}}(f)|$}  & \multirow{ 2}{*}{\bf \# Overlaps} \\ \cline{5-8}
    & & & &\bf Mean & \bf SD & \bf Mean & \bf SD & \\ \hline
    David (Fig.~\ref{fig:edem_area}(a)) & 21338 & 2.6982
&(1,1.2,1.4) & 0.7078 & 0.5248 & 0.0132 & 0.0136 & 0 \\  \hline
    Buddha (Fig.~\ref{fig:edem_area}(b)) & 50002  & 10.5964
 &(1,1,1.4) & 0.4365 &  0.3907 &  0.0366 & 0.0648 & 0 \\ \hline 
    Hippocampus (Fig.~\ref{fig:edem_area}(c)) & 12000 & 3.1002
 & (1,1,2.2) & 1.3635 & 1.4740 & 0.0515 & 0.0973  & 0 \\ \hline 
    Twisted Ball (Fig.~\ref{fig:edem_area}(d)) & 38620 & 7.5060
 & (1,0.8,1.4) & 0.8998 & 0.8195 & 0.0239 & 0.0289 & 0 \\ \hline 
    Chinese Lion 1 (Fig.~\ref{fig:front_add}(a)) & 10000 & 10.1311
 & (1,1,0.8) & 1.8306 & 1.8851 & 0.1407 & 0.2081 & 0  \\ \hline
    Chinese Lion 2 (Fig.~\ref{fig:front_add}(a)) & 10000 & 9.4919
 & (1,1,1.1) & 1.6529 & 1.7923 & 0.1180 & 0.1759 & 0  \\ \hline
    Chinese Lion 3 (Fig.~\ref{fig:front_add}(a)) & 10000 & 7.9418
 & (1,1,1.4) & 1.5394 & 1.8125 & 0.1137 & 0.1892 & 0  \\ \hline
  \end{tabular}
}
\end{center}
\end{table}

\begin{table}[t!]
\small
    \caption{\textbf{The results of the Hippocampus model obtained by our EDEM algorithm with different elliptic radii.} Here, for different choices of the elliptic radii $(a,b,c)$ of the target ellipsoid, we record the mean and standard deviation of the final log-area error $|d_{\text{area}}(f)|$, the mean value of the norm of the Beltrami coefficient $|\mu|$, and the number of overlaps.}\label{tab:EDEM_radii} 
  \begin{center}
  \begin{tabular}{|c|c|c|c|c|c|c|c|} \hline 
    $\mathbf{a}$ & $\mathbf{b}$ & $\mathbf{c}$
      & \bf $\text{Mean}(|d_{\text{area}}(f)|)$   & \bf $\text{SD}(|d_{\text{area}}(f)|)$  & \bf Mean$(|\mu|)$  & \bf \# Overlaps  \\\hline 
    \multirow{10}{*}{1} & 1 & 1 & 0.0895 & 0.1684 & 0.4443 & 0  \\ \cline{2-7} 
     & 0.7 & 1 & 0.1349 & 0.3481 & 0.4620 & 0 \\ \cline{2-7} 
     & 1.2 & 1 & 0.1042 & 0.2063 & 0.4446 & 0 \\ \cline{2-7} 
     & 1 & 0.8 & 0.0996 & 0.1838 & 0.4709 & 0 \\ \cline{2-7} 
     & 1 & 1.3 & 0.0816 & 0.1518 & 0.4017 & 0 \\ \cline{2-7} 
     & 1 & 1.5 & 0.0730 & 0.1322 & 0.3730 & 0 \\ \cline{2-7} 
     & 1 & 2.2 & 0.0515 & 0.0973 & 0.2862 & 0  \\ \cline{2-7} 
     & 1.3 & 1.6 & 0.1484 & 0.3474 & 0.4026 & 0 \\ \cline{2-7} 
     & 1.2 & 1.8 & 0.1340 & 0.3179 & 0.3942 & 0 \\ \cline{2-7} 
     & 1.1 & 2 & 0.0975 & 0.2178 & 0.3718 & 0 \\ \hline 
  \end{tabular}
\end{center}
\end{table}

It is natural to ask whether the performance of our EDEM method is affected by the shape of the target ellipsoid. Here, we map the Hippocampus model to various target ellipsoids with different elliptic radii using our EDEM method and analyze the area and angle distortions of the resulting mappings. In Table~\ref{tab:EDEM_radii}, we report the mean and variance of the log-area error, as well as the mean value of the norm of the Beltrami coefficient $|\mu|$. It can be observed that for all combinations of the elliptic radii, all mapping results are folding-free and possess low log-area error. This shows that our EDEM method is capable of computing area-preserving parameterizations of genus-0 closed surfaces onto different prescribed ellipsoids. However, if we also take the angle distortion into consideration, then some differences can be observed. For instance, if we increase the value of $c$, we can see that the angle distortion is further reduced. This experiment suggests that the overall ellipsoidal shape plays an important role and motivates the need for the proposed EDEQ method, which optimizes the ellipsoidal geometry and achieves ellipsoidal density-equalizing quasi-conformal maps.

\begin{figure}[t!]
    \centering
    \includegraphics[width=\textwidth]{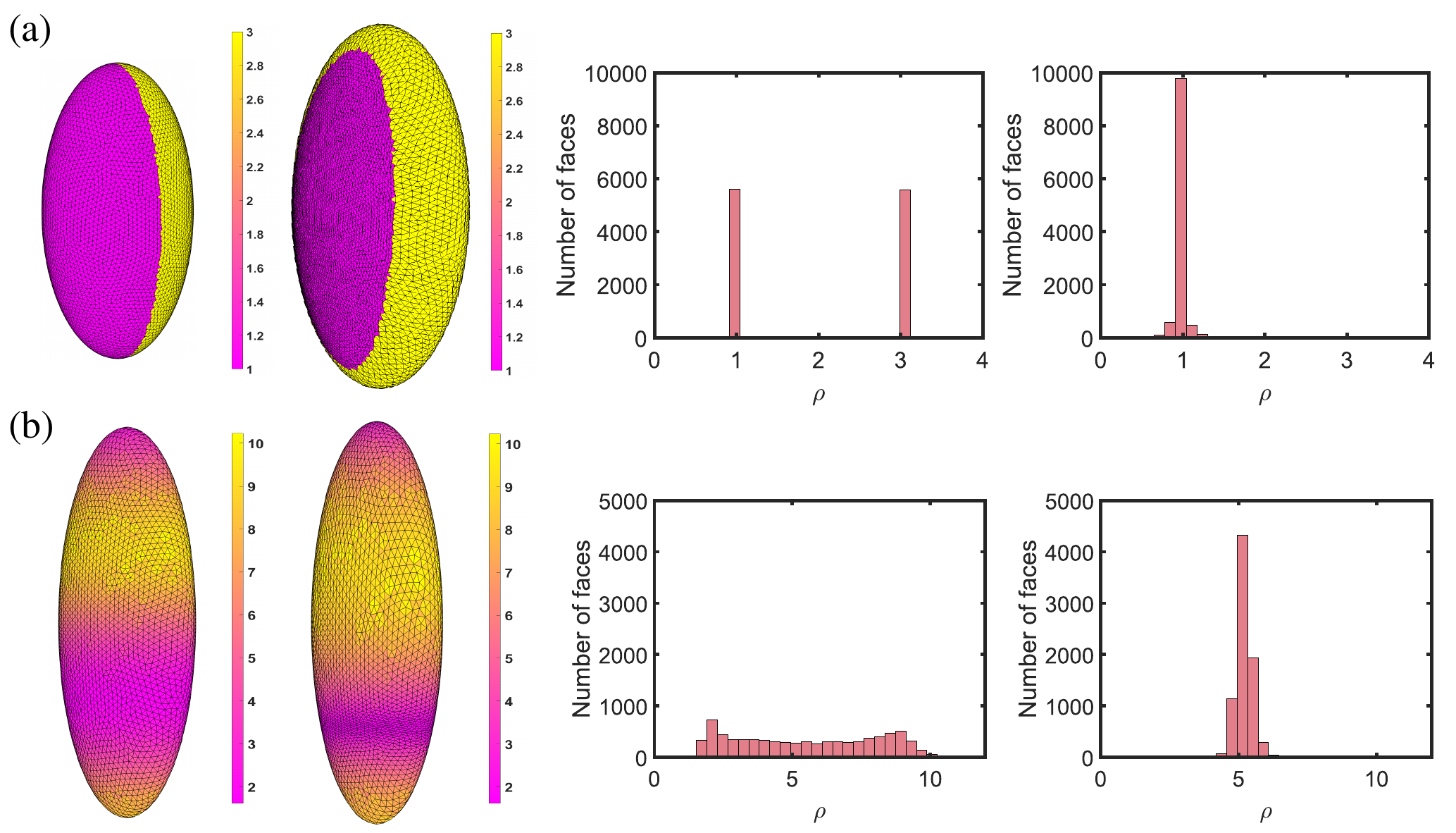}
    \caption{\textbf{Ellipsoidal density-equalizing quasi-conformal maps of ellipsoidal surfaces.} Each row shows one example. (a)~An example with discontinuous input density. Here, the initial elliptic radii are $(a_0,b_0,c_0) = (1,1,2)$ and the final elliptic radii are $(a,b,c) = (1,1.8973,2.8163)$. (b)~An example with continuous input density. Here, the initial elliptic radii are $(a_0,b_0,c_0) = (1,2,4)$ and the final elliptic radii are $(a,b,c) = (1,1.9888,4.3110)$. Left to right: The initial ellipsoidal surface color-coded with the initial density, the final EDEQ result color-coded with the initial density, the histogram of the initial density, and the histogram of the final density.}
    \label{fig:edeq_sythetic}
\end{figure}

\subsection{Ellipsoidal density-equalizing quasi-conformal map}
Next, we test our EDEQ algorithm for computing the ellipsoidal density-equalizing quasi-conformal mappings. We set $\alpha = 5$ in the following experiments. In Fig.~\ref{fig:edeq_sythetic}(a), we define different populations at different regions on the ellipsoid to obtain a discontinuous initial density. Note that the maximum density is three times larger than the minimum one. It can be observed in the EDEQ mapping result that the shape of the ellipsoid is changed under the mapping process, while the higher-density domain expands and the lower-density domain contracts. Comparing the initial and final density histograms, it can be observed that the density is highly equalized. Fig.~\ref{fig:edeq_sythetic}(b) shows another example with a continuous initial density defined on an ellipsoid. Under the EDEQ algorithm, the ellipsoidal shape is stretched along the z-direction and becomes more elongated, and different regions are expanded or shrunk according to the given density. From the density histograms, it can again be observed that the mapping result is highly density-equalizing.

\begin{figure}[t!]
    \centering
    \includegraphics[width=\textwidth]{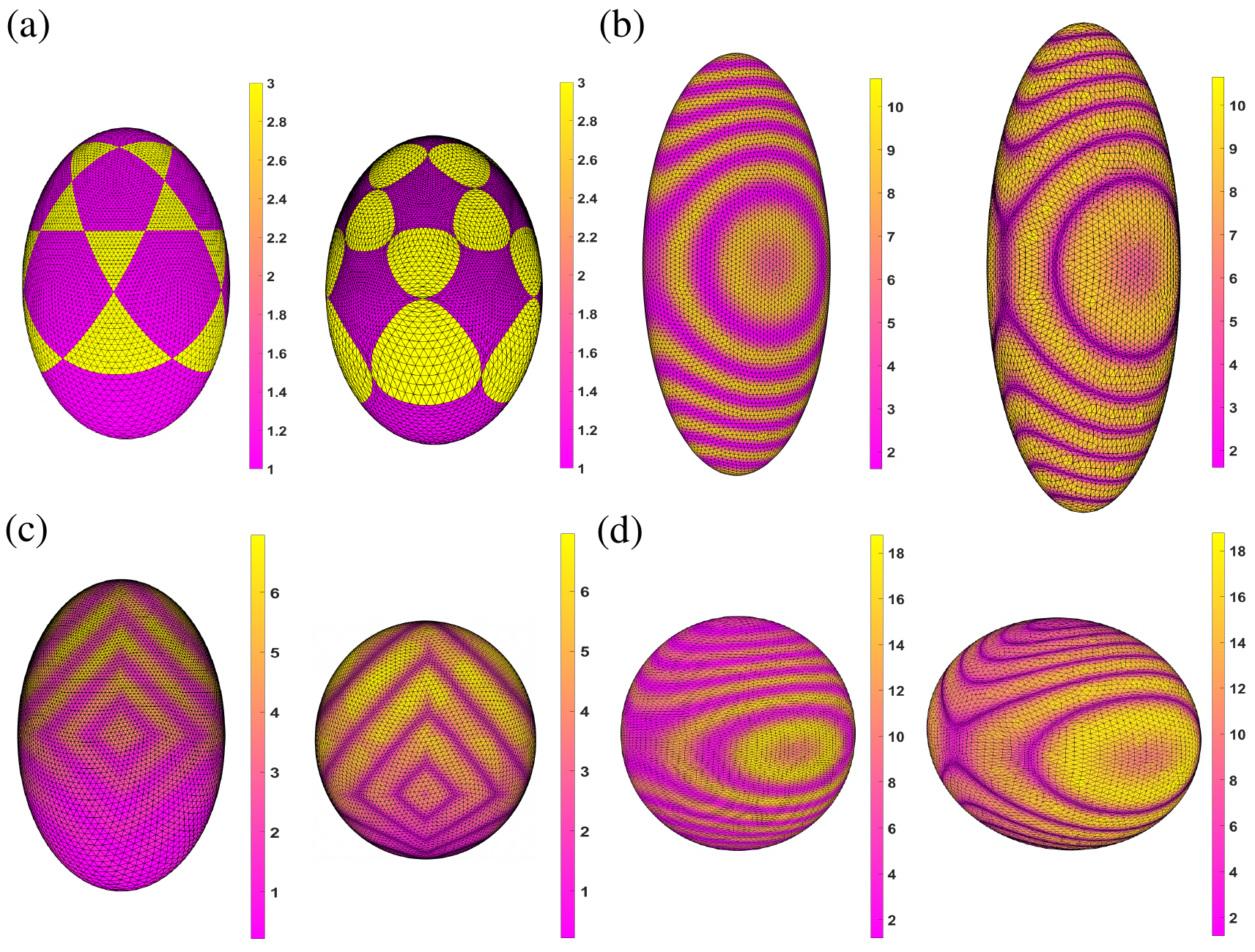}
    \caption{\textbf{Ellipsoidal density-equalizing quasi-conformal maps of ellipsoidal surfaces.} For each example, the left figure shows the input surface color-coded with the initial density, and the right figure shows the EDEQ result color-coded with the initial density. (a)~An ellipsoid with non-uniformly distributed mesh elements and a prescribed discontinuous density. Here, the initial elliptic radii are $(a_0,b_0,c_0) = (1,1,1.5)$ and the final elliptic radii are $(a,b,c) = (1,1,1.4250)$. (b)~An elongated ellipsoid with a prescribed continuous density. Here, the initial elliptic radii are $(a_0,b_0,c_0) = (1,0.6,2)$ and the final elliptic radii are $(a,b,c) = (1,0.9821,2.5162)$. (c)~An ellipsoid with non-uniformly distributed mesh elements and a complex prescribed density. Here, the initial elliptic radii are $(a_0,b_0,c_0) = (1,1,1.5)$ and the final elliptic radii are $(a,b,c) = (1,1.0310,1.0783)$. (d)~A sphere with a prescribed continuous density. Here, the initial elliptic radii are $(a_0,b_0,c_0) = (1,1,1)$ and the final elliptic radii are $(a,b,c) = (1,1.3095,0.9100)$. }
    \label{fig:sythetic_edeq_add}
\end{figure}

In Fig.~\ref{fig:sythetic_edeq_add}, we further apply the EDEQ method to the same set of ellipsoidal examples in Fig.~\ref{fig:sythetic_edem_add}. It can be observed that our EDEQ method produces ellipsoidal density-equalizing mapping results with both the ellipsoidal geometry and the vertex positions optimized. For a more quantitative analysis, we record the initial and final elliptic radii, the variance of the initial and final densities, the mean value of the Beltrami coefficient $\mu$, and the number of overlaps for all the above-mentioned examples in Table~\ref{tab:EDEQ}. From the table, we can see that the ellipsoidal shapes are effectively optimized under the EDEQ method. Furthermore, in addition to possessing a highly density-equalizing effect as reflected in the variance of the final density, the mapping results also achieve a low conformal distortion. Specifically, note that the average values of the norm of the Beltrami coefficient $\text{mean}(|\mu|)$ in all examples are around 0.2 only, while for general bijective quasi-conformal maps, the range of $|\mu|$ is from 0 to 1, with $|\mu| = 0$ indicating that the mapping is conformal and $|\mu| = 1$ indicating that there is a singularity. Moreover, the $|\mu|$ value of our method is significantly smaller than that of fixed shape methods, indicating that our method can better preserve the local geometric structure. Hence, we can see that the EDEQ method not only achieves density equalization but also reduces the conformal distortion of the ellipsoidal mappings.

\begin{table}[t!]
\small
    \caption{\textbf{The performance of our proposed EDEQ algorithm.} For each surface, we record the number of triangle elements, the initial elliptic radii $(a_0,b_0,c_0)$ of the ellipsoid, the final elliptic radii $(a,b,c)$ of the ellipsoidal shape obtained by EDEQ, the variance of the normalized initial density $\widetilde{\rho}_1 = \frac{\rho_1}{\text{Mean}(\rho_1)}$ (where $\rho_1$ is the initial vertex density), the variance of the normalized final density $\widetilde{\rho}_2  = \frac{\rho_2}{\text{Mean}(\rho_2)}$ (where $\rho_2$ is the final vertex density), the mean value of norm of the Beltrami coefficient $|\mu|$, and the number of overlaps.}\label{tab:EDEQ} 
  \begin{center}
\resizebox{\textwidth}{!}{
  \begin{tabular}{|C{16mm}|c|c|c|c|c|c|c|} \hline
    \bf Surface & \bf \# Faces & \bf $\mathbf{(a_0,b_0,c_0)}$ & \bf $\mathbf{(a,b,c)}$ &\bf $\text{Var}(\widetilde{\rho}_1)$  &\bf $\text{Var}(\widetilde{\rho}_2)$ & \bf \text{mean}($|\mu|$)  & \bf \# Overlaps \\\hline
    Ellipsoid 1 (Fig.~\ref{fig:edeq_sythetic}(a)) & 7808 & (1,1,2) & (1,1.8973,2.8163)  & 0.2503 & 0.0049 & 0.1434 & 0  \\ \hline
    Ellipsoid 2 (Fig.~\ref{fig:edeq_sythetic}(b)) & 7808 & (1,2,4) & (1,1.9888,4.3110)  & 0.2084 & 0.0029 & 0.1953 & 0   \\ \hline
    Ellipsoid 3 (Fig.~\ref{fig:sythetic_edeq_add}(a)) & 20480 & (1,1,1.5) & (1,1,1.4250) & 0.333 & 0.0153 & 0.1852 & 0  \\ \hline
    Ellipsoid 4 (Fig.~\ref{fig:sythetic_edeq_add}(b)) & 18904 & (1,0.6,2) & (1,0.9821,2.5162)  & 0.2135 & 0.0082 & 0.2285 & 0 \\ \hline
    Ellipsoid 5 (Fig.~\ref{fig:sythetic_edeq_add}(c)) & 20480 & (1,1,1.5) & (1,1.0310,1.0783) & 0.2580 & 0.0011 & 0.1679 & 0  \\ \hline
    Ellipsoid 6 (Fig.~\ref{fig:sythetic_edeq_add}(d)) & 18904 & (1,1,1) & (1,1.3095,0.9100) & 0.2940 & 0.0072 & 0.2267 & 0 \\ \hline
  \end{tabular}
  }
\end{center}
\end{table}

\begin{figure}[!h]
    \centering
    \includegraphics[width=\textwidth]{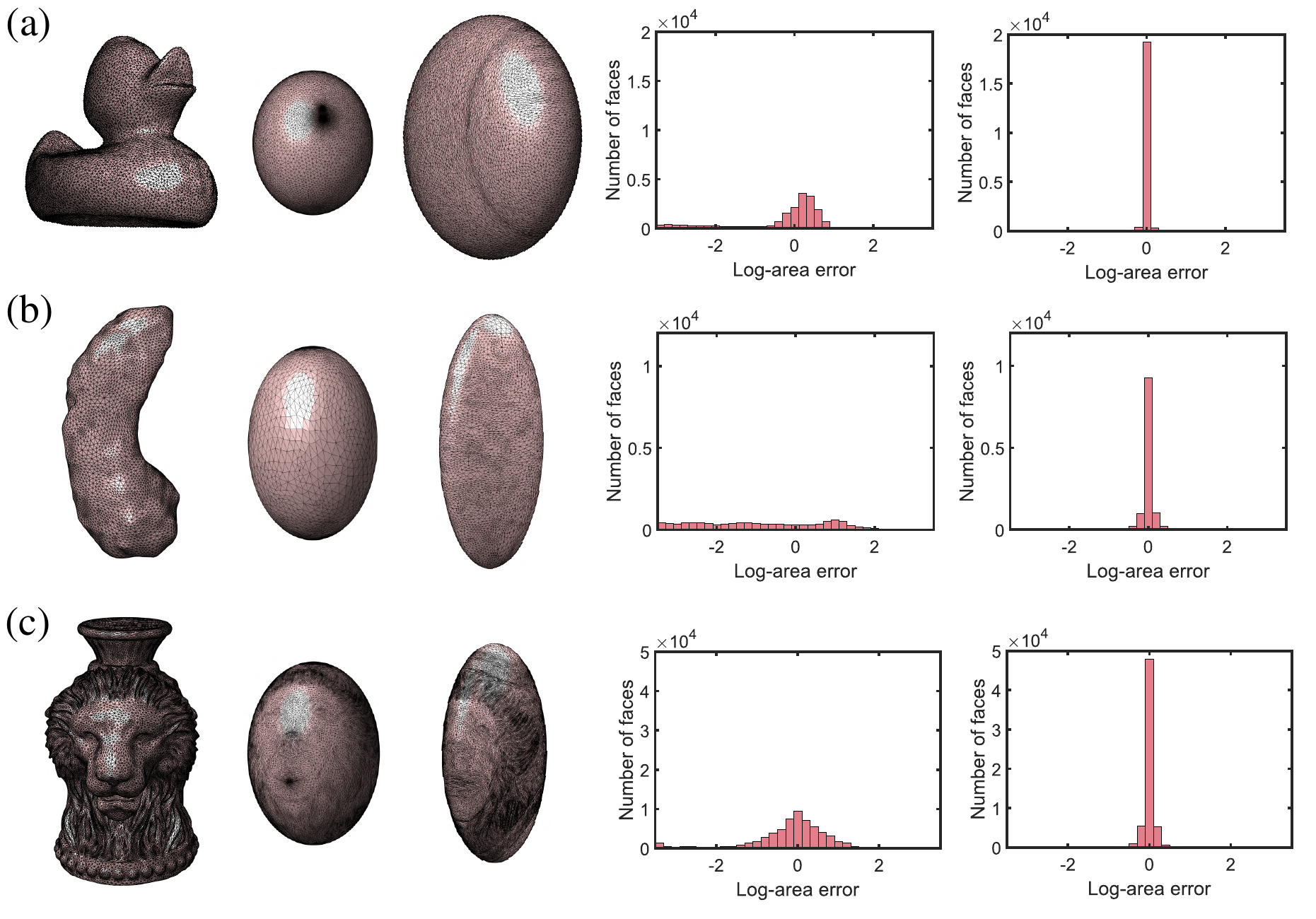}
    \caption{\textbf{Ellipsoidal area-preserving quasi-conformal parameterization of genus-0 closed surfaces.} Each row shows one example. (a)~The Duck model with initial elliptic radii $(a_0,b_0,c_0) = (1,1,1.2)$ and final elliptic radii $(a,b,c) = (1,1.8797,1.9511)$. (b)~The Hippocampus model with initial elliptic radii $(a_0,b_0,c_0) = (1,1,1.5)$ and final elliptic radii $(a,b,c) = (1,0.4487,2.0513)$. (c)~The Lion-Vase model with initial elliptic radii $(a_0,b_0,c_0) = (1,1,1.4)$ and final elliptic radii $(a,b,c) = (1,0.47557,1.6682)$. Left to right: The input surface mesh, the initial ellipsoidal conformal parameterization, the final EDEQ result, the histogram of the log-area error $d_{\text{area}}$ of the initial ellipsoidal parameterization, and the histogram of the log-area error $d_{\text{area}}$ of the final ellipsoidal parameterization.}
    \label{fig:edeq_ellipsoid}
\end{figure}

Next, we consider using the EDEQ method to compute the ellipsoidal area-preserving parameterizations for genus-0 closed surfaces. For each surface in Fig.~\ref{fig:edeq_ellipsoid}, we first conformally map it onto an ellipsoid with some arbitrary elliptic radii using the FECM method~\cite{choi2024fast}. Then, we apply the EDEQ algorithm with the population set as the original face area to obtain the ellipsoidal area-preserving parameterizations. It can be observed that the ellipsoidal shapes are changed significantly under the EDEQ method. Comparing the initial and final log-area error histograms, we can see that the mapping results are highly area-preserving. 

\begin{figure}[!h]
    \centering
    \includegraphics[width=\textwidth]{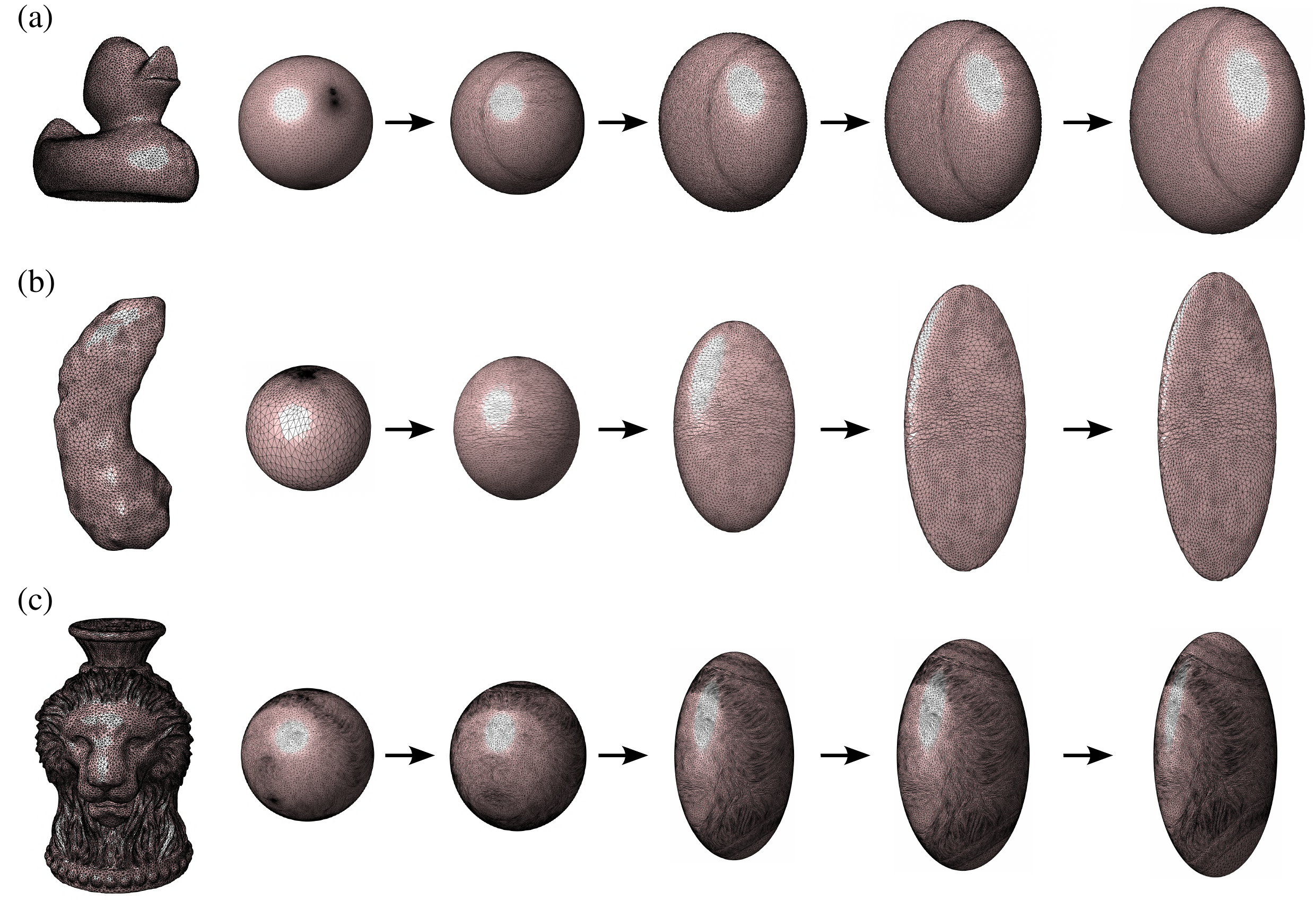}
    \caption{\textbf{The shape change of the ellipsoidal parameterization under the proposed EDEQ method.} Each row shows one example. (a)~The Duck model. (b)~The Hippocampus model. (c)~The Lion-Vase model. Left to right: The input surface mesh, the initial ellipsoidal conformal parameterization, the results after 10, 50, and 100 iterations, and the final EDEQ result.}
    \label{fig:edeq_area}
\end{figure}

As the choice of the initial elliptic radii is arbitrary, we can even start with a unit sphere and run the EDEQ method to eventually obtain an optimal ellipsoidal parameterization. In Fig.~\ref{fig:edeq_area}, we consider several genus-0 closed surface models and compute an initial spherical conformal parameterization. Then, we apply the proposed EDEQ algorithm to compute an optimal ellipsoidal area-preserving parameterization. It can be observed that throughout the EDEQ iterative process, both the ellipsoidal shapes and the vertex positions on the ellipsoids are changed gradually. For a more quantitative analysis, Table~\ref{tab:EDEQ_area} summarizes the performance of our EDEQ algorithm in terms of the final radii, the initial and final log-area error, the conformal distortion, and the number of overlaps. Comparing the initial and final log-area error, it can be observed that the log-area error is significantly reduced in all examples. The mean of the norm of the Beltrami coefficient $|\mu|$ is also small, indicating that the conformal distortion is low. Besides, the number of overlaps is 0 for all examples, which shows that our mapping results are all folding-free. In Table~\ref{tab:comparision}, we further compare the performance of the EDEQ method and the EDEM method in terms of the area and conformal distortions, where both methods use the same initial elliptic radii. It can be observed that the EDEQ method achieves a significantly lower conformal distortion when compared to the EDEM method, while the log-area error remains comparable. Besides, we remark that since the EDEQ mappings are diffeomorphisms, they preserve the topological structure between the input genus-0 closed surface and the final ellipsoidal domain. Therefore, the final results of EDEQ will still be genus-0 closed ellipsoidal surfaces with $a,b,c >0$ and will not become a flat disk. In our experiments, we empirically tested convergence and observed the radii to asymptotically approach non-zero values. Altogether, the above experiments demonstrate the effectiveness of our proposed EDEQ method for ellipsoidal density-equalizing quasi-conformal maps.

\begin{table}[t!]
\small
    \caption{\textbf{The performance of our EDEQ algorithm for ellipsoidal area-preserving parameterization.} For each surface, we start with an initial spherical conformal parameterization and then apply the EDEQ algorithm. The number of triangle elements, the final elliptic radii $(a,b,c)$, the mean and standard deviation of the initial log-area error $|d_{\text{area}}(f_0)|$ and the final log-area error $|d_{\text{area}}(f)|$, the mean value of the norm of the Beltrami coefficient $|\mu|$, and the number of overlaps are recorded.}\label{tab:EDEQ_area}
    
  \begin{center}
\resizebox{\textwidth}{!}{
  \begin{tabular}{|C{20mm}|c|c|c|c|c|c|c|c|} \hline
    \multirow{ 2}{*}{\bf Surface} & \multirow{ 2}{*}{\bf \# Faces} & \multirow{ 2}{*}{\bf $\mathbf{(a,b,c)}$}  & \multicolumn{2}{c|}{\bf $|d_{\text{area}}(f_0)|$} & \multicolumn{2}{c|}{\bf $|d_{\text{area}}(f)|$} & \multirow{ 2}{*}{\bf \text{mean}($|\mu|$)} & \multirow{ 2}{*}{\bf \# Overlaps}  \\ \cline{4-7}
    & & &\bf Mean & \bf SD & \bf Mean & \bf SD & & \\ \hline
    Duck (Fig.~\ref{fig:edeq_area}(a)) & 20000 & (1,1.8797,1.8521) & 1.0561 & 1.2190 & 0.0241 & 0.0379 & 0.2216 &  0 \\  \hline
    Hippocampus (Fig.~\ref{fig:edeq_area}(b)) & 12000 & (1,0.1886,1.8818)  & 2.5651 & 1.7272 & 0.1028 & 0.1990 & 0.2602 &   0 \\ \hline 
    Lion-Vase (Fig.~\ref{fig:edeq_area}(c)) & 38620 & (1,0.3833,1.6136)  & 0.8264 & 0.7195 & 0.1032 & 0.1749 & 0.2306 &  0 \\ \hline 
    David (Fig.~\ref{fig:front_add}(b)) & 21338 & (1,1.0062,1.5126)  & 0.5652 & 0.4349 & 0.0125 & 0.0133 & 0.1279  & 0 \\ \hline
  \end{tabular}
}
\end{center}
\end{table}

\begin{table}[!h]
  \caption{\textbf{Comparison between the EDEM method and the EDEQ method.} We record the mean and standard deviation of the final area distortion $|d_{\text{area}}(f)|$, the mean of the norm of the Beltrami coefficient $|\mu|$, and the number of overlaps for each example and each method.} \label{tab:comparision}
\begin{center}
\resizebox{\textwidth}{!}{
\begin{tabular}{|c|C{17mm}|C{17mm}|c|c|C{17mm}|C{17mm}|c|c|}
\hline
\bf Surface & \multicolumn{4}{c|}{\textbf{EDEM}}&\multicolumn{4}{c|}{\textbf{EDEQ}} \\ 
\cline{2-9}
\multicolumn{1}{|c|}{}& \bf Mean of $|d_{\text{area}}(f)|$ & \bf SD of $|d_{\text{area}}(f)|$  & \bf Mean$(|\mu|)$ &\bf \# Overlaps &\bf Mean of $|d_{\text{area}}(f)|$ & \bf SD of $|d_{\text{area}}(f)|$  & \bf Mean$(|\mu|)$ & \bf \# Overlaps\\ \hline
Duck  & 0.0268 & 0.0507 & 0.2467 & 0 & 0.0241 & 0.0379 & 0.2216 & 0 \\ \hline
Hippocampus & 0.0895  & 0.1684 & 0.4443 & 0 & 0.1028 & 0.1990 & 0.2602 & 0  \\ \hline
Lion-Vase  & 0.0909  & 0.1604 & 0.3137  & 0 & 0.1032 & 0.1749 & 0.2306  & 0 \\\hline
\end{tabular}
}
\end{center}
\end{table}

\section{Application to genus-0 surface remeshing}\label{sec:applications}
Note that one important application of surface parameterization is surface remeshing~\cite{alliez2008recent}. Specifically, after computing the parameterization of a given surface, one can generate a high-quality mesh on the parameter domain and use the inverse of the parameterization mapping to map the mesh onto the input surface, thereby yielding a remeshed surface. For instance, Sheffer and de Sturler~\cite{sheffer2001parameterization} utilized the angle-based flattening method to compute parameterizations for surface remeshing. Praun and Hoppe~\cite{praun2003spherical} computed spherical parameterization for surface remeshing. Later, different point cloud parameterization methods~\cite{zwicker2004meshing,choi2016spherical} have been developed and applied to surface meshing. Analogously, our proposed EDEM and EDEQ methods can be easily applied to genus-0 surface remeshing. 

\begin{figure}[t!]
    \centering
    \includegraphics[width=\textwidth]{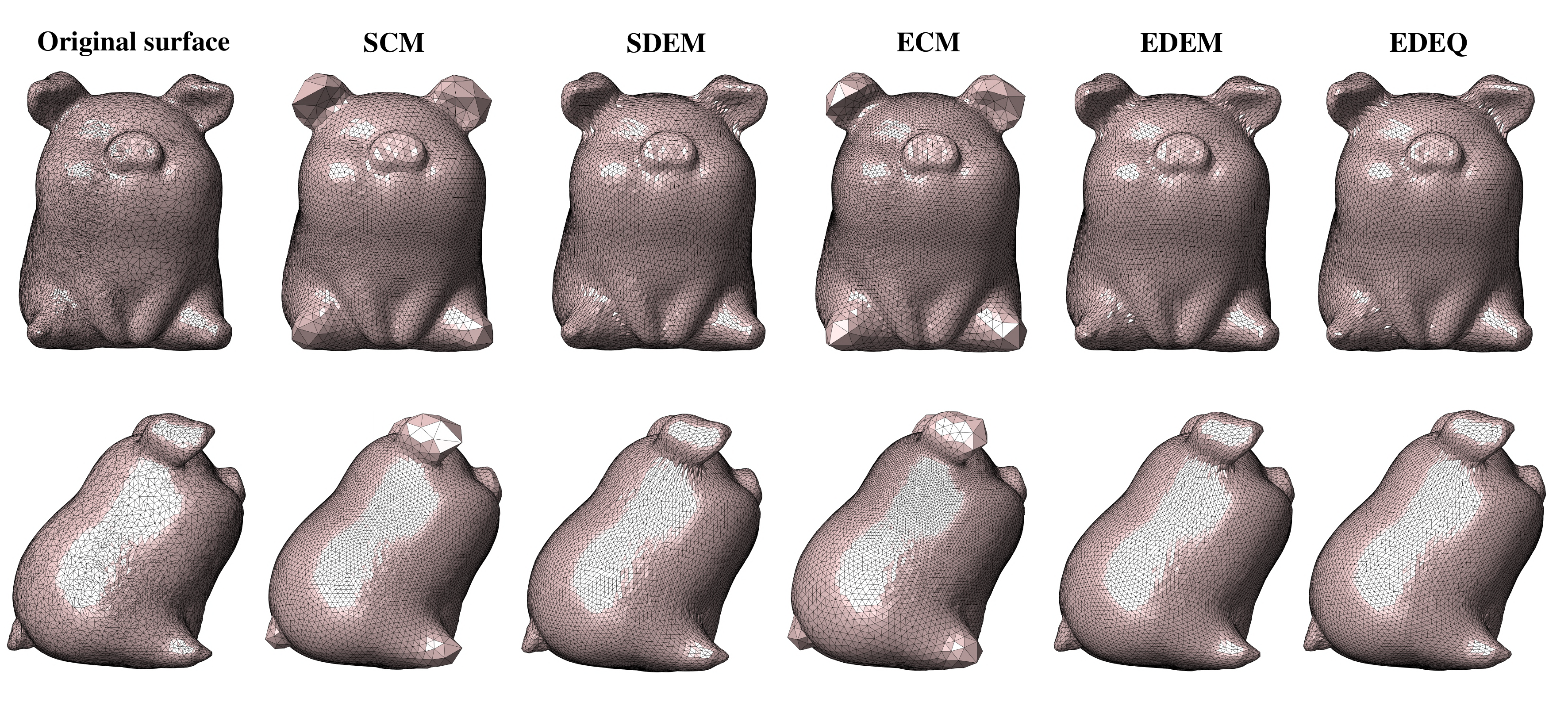}
    \caption{\textbf{The surface remeshing results of the Pig model obtained by SCM~\cite{choi2015flash}, SDEM~\cite{lyu2024spherical}, ECM~\cite{choi2024fast}, the proposed EDEM method, and the proposed EDEQ method.} For each surface, two different views are provided. }
    \label{fig:remeshing_pig}
\end{figure}

\begin{figure}[t!]
    \centering
    \includegraphics[width=\textwidth]{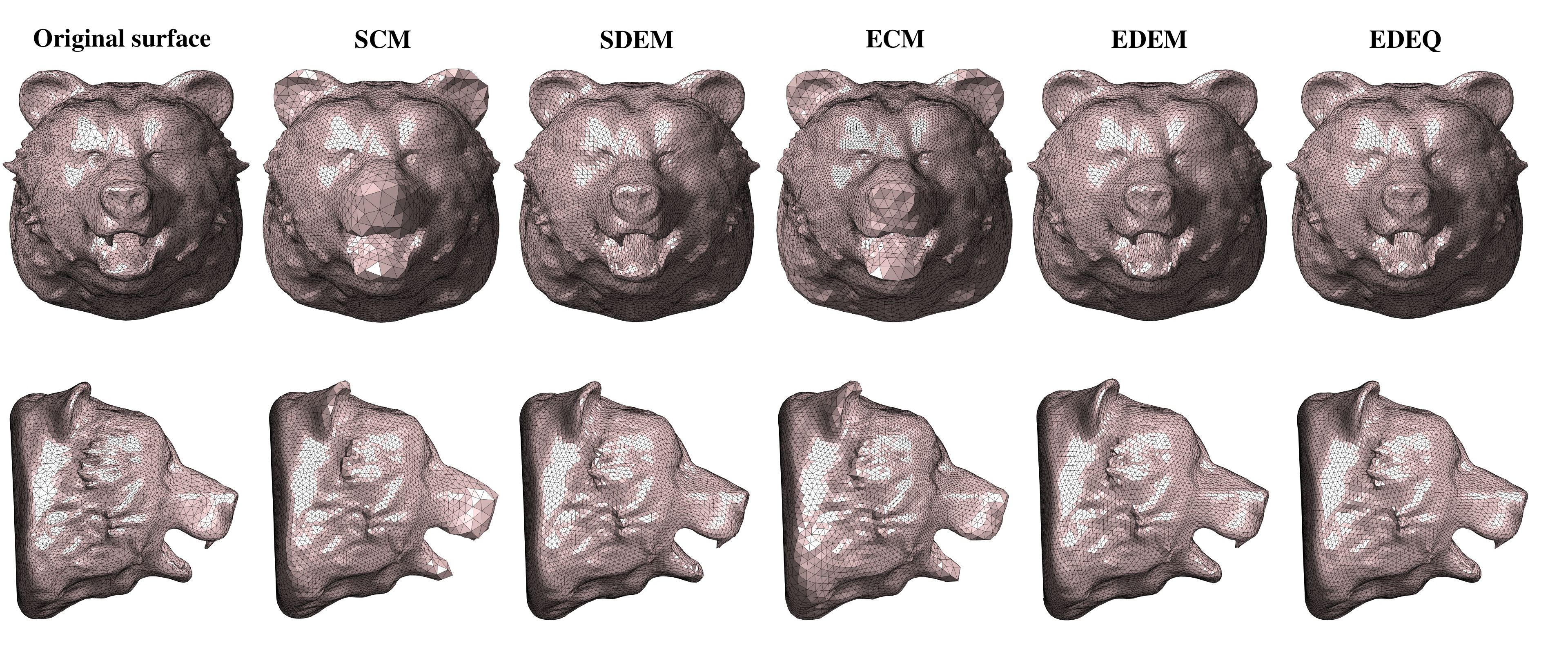}
    \caption{\textbf{The surface remeshing results of the Bear model obtained by SCM~\cite{choi2015flash}, SDEM~\cite{lyu2024spherical}, ECM~\cite{choi2024fast}, the proposed EDEM method, and the proposed EDEQ method.} For each surface, two different views are provided. }
    \label{fig:remeshing_bear}
\end{figure}

\begin{figure}[t!]
    \centering
    \includegraphics[width=\textwidth]{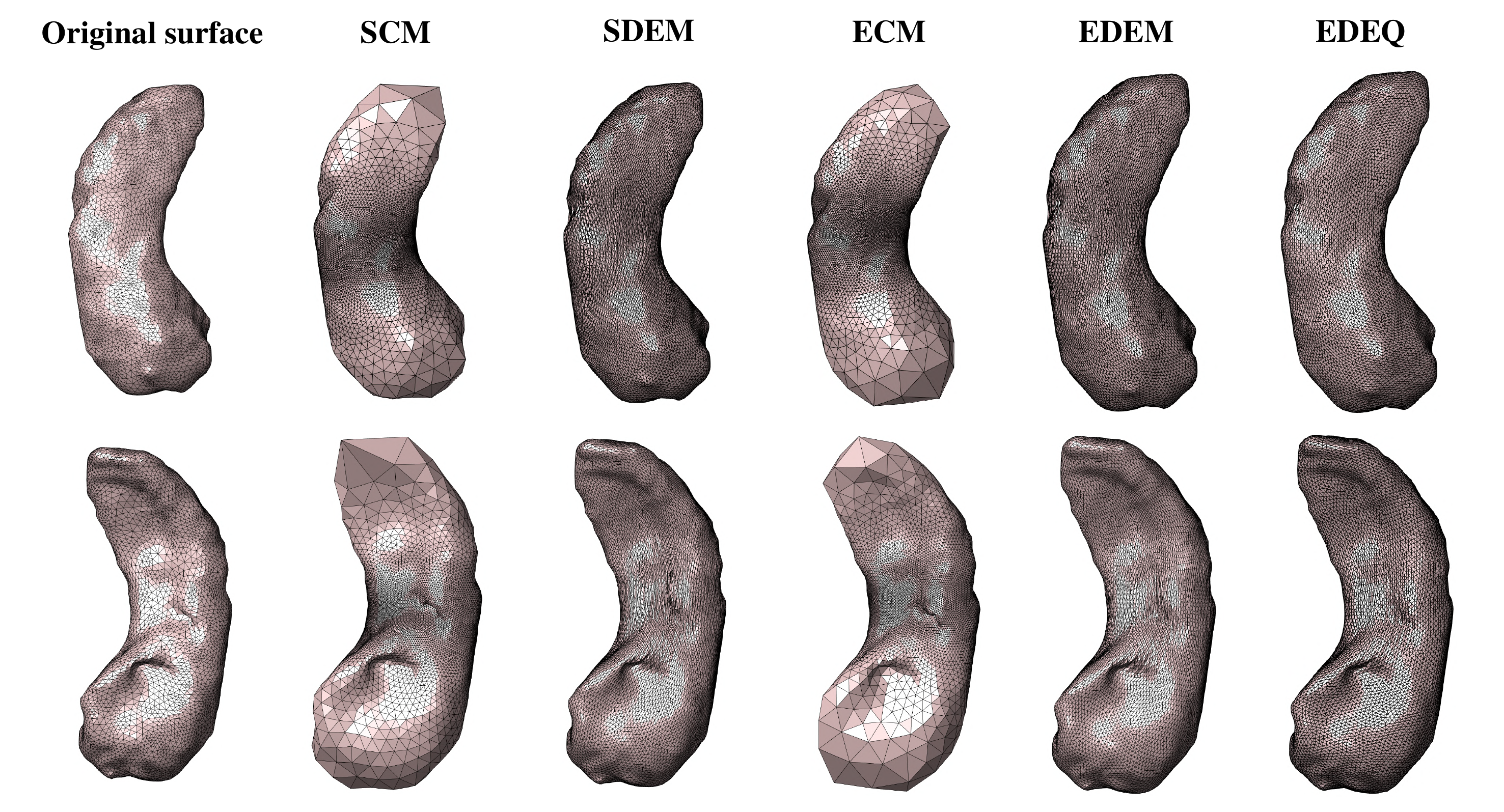}
    \caption{\textbf{The surface remeshing results of the Hippocampus model obtained by SCM~\cite{choi2015flash}, SDEM~\cite{lyu2024spherical}, ECM~\cite{choi2024fast}, the proposed EDEM method, and the proposed EDEQ method.} For each surface, two different views are provided. }
    \label{fig:remeshing_hippocampus}
\end{figure}

\begin{figure}[t!]
    \centering
    \includegraphics[width=\textwidth]{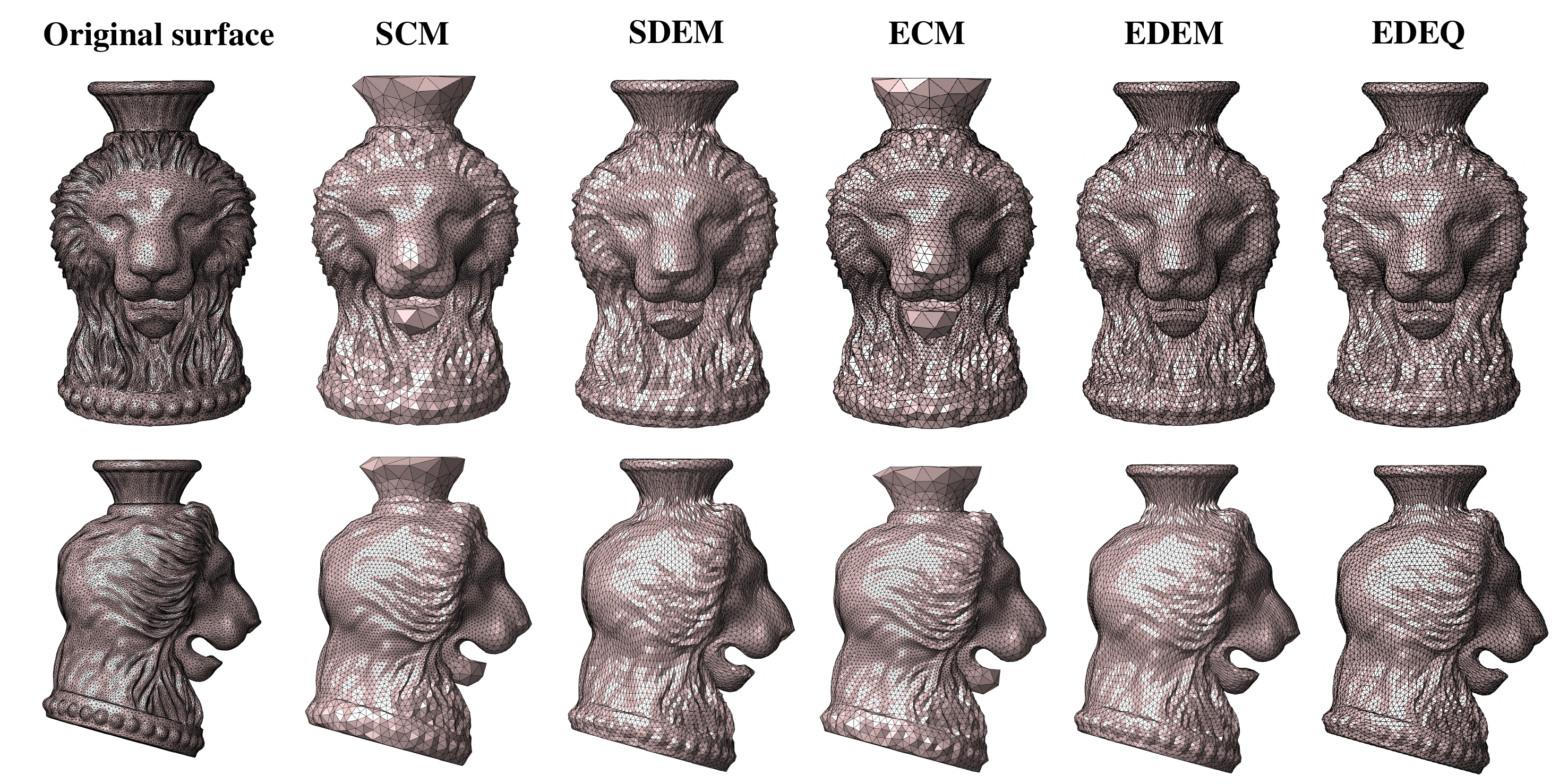}
    \caption{\textbf{The surface remeshing results of the Lion-vase model obtained by SCM~\cite{choi2015flash}, SDEM~\cite{lyu2024spherical}, ECM~\cite{choi2024fast}, the proposed EDEM method, and the proposed EDEQ method.} For each surface, two different views are provided. }
    \label{fig:remeshing_vaselion}
\end{figure}

In Fig.~\ref{fig:remeshing_pig}, Fig.~\ref{fig:remeshing_bear}, Fig.~\ref{fig:remeshing_hippocampus}, and Fig.~\ref{fig:remeshing_vaselion}, we consider various genus-0 surface models and compare the remeshing results obtained by different parameterization methods for genus-0 closed surfaces, including spherical conformal mapping (SCM)~\cite{choi2015flash}, spherical density-equalizing mapping (SDEM)~\cite{lyu2024spherical}, ellipsoidal conformal mapping (ECM)~\cite{choi2024fast}, our ellipsoidal density-equalizing mapping (EDEM) method, and our ellipsoidal density-equalizing quasi-conformal mapping (EDEQ) method. For each genus-0 surface model $\mathcal{M}$ in $\mathbb R^3$ and each parameterization method, denote the parameterization mapping by $f: \mathcal{M} \rightarrow \mathcal{T}$, where the parameter domain $\mathcal{T}$ is either the unit sphere (for SCM and SDEM) or an ellipsoid (for ECM, EDEM and EDEQ). Then we construct a regular triangular mesh $\mathcal{T}_{\Delta}$ on $\mathcal{T}$ using the DistMesh method~\cite{persson2004simple}. Finally, using the inverse mappings $f^{-1}: \mathcal{T} \to \mathcal{M}$, the regular mesh $\mathcal{T}_{\Delta}$ can be mapped back onto $\mathcal{M}$, yielding a remeshed surface $f^{-1}(\mathcal{T}_{\Delta})$. For a fair comparison, we enforce the number of vertices of the triangular mesh generated by DistMesh (and hence the number of vertices of the final remeshing result) for each method to be approximately 8500 (with variation $< \pm 1\%$).

From the remeshing results in Fig.~\ref{fig:remeshing_pig},  Fig.~\ref{fig:remeshing_bear}, Fig.~\ref{fig:remeshing_hippocampus}, and Fig.~\ref{fig:remeshing_vaselion}, it can be observed that the two conformal approaches (SCM and ECM) lead to significantly non-uniform remeshing results. Moreover, the shape features near the sharp regions are lost in the SCM and ECM results. This can be explained by the fact that the conformal mappings preserve angles but may yield large area distortions. By contrast, the remeshing results obtained by the density-equalizing approaches (SDEM, EDEM, and EDEQ) are much more uniform and have a higher mesh quality. The features located near the sharp regions are well-preserved.   

\begin{figure}[t!]
    \centering
    \includegraphics[width=\textwidth]{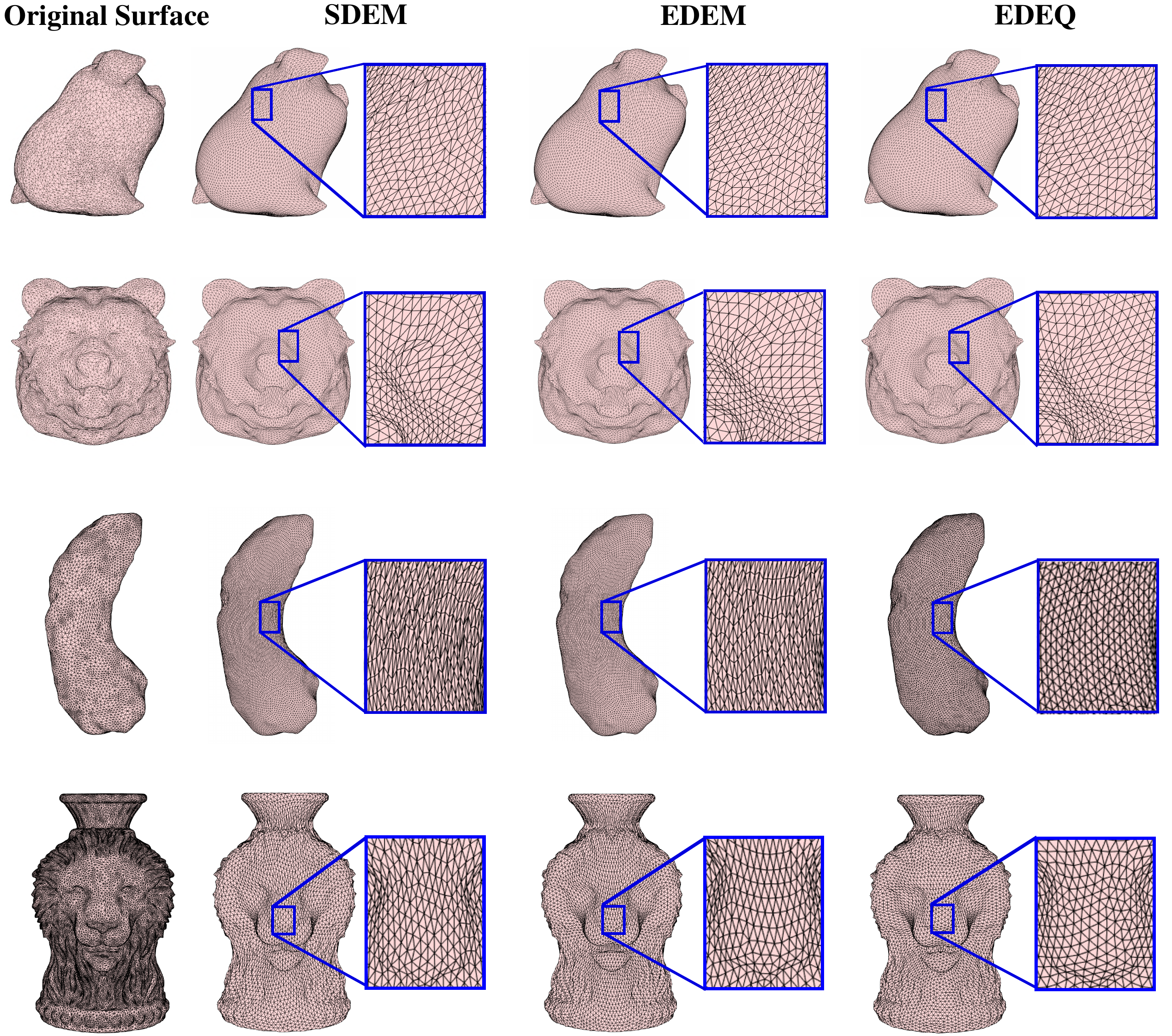}
    \caption{\textbf{Comparison for the surface remeshing results between the SDEM method~\cite{lyu2024spherical}, the proposed EDEM method, and the proposed EDEQ method.} Each row shows one example. For each method, we show a zoom-in of the remeshed surface obtained by the method to visualize the triangle quality.}
    \label{fig:remeshing_detail} 
\end{figure}

To further compare the remeshing results obtained using the density-equalizing approaches, Fig.~\ref{fig:remeshing_detail} shows the zoom-in images of the SDEM, EDEM and EDEQ remeshing results for the Pig, Bear, Hippocampus, and Lion-Vase models. From the zoom-in images, we can see that the EDEQ remeshing results are more uniform than the ones by the other two methods, especially at the regions corresponding to certain sharp features in the surface models. Moreover, comparing the mesh qualities in terms of the angles of the triangles, it can be observed that the SDEM remeshing results contain more skinny and irregular triangle elements when compared to the other two methods. This can be explained by the fact that the SDEM method maps all surfaces onto the unit sphere regardless of their geometries, which may lead to large geometric distortions. Also, while the EDEM method allows us to map the surfaces onto a prescribed ellipsoid, we may need to carefully choose a suitable ellipsoidal shape in order to reduce both the area and angle distortions of the mappings. By contrast, the EDEQ method optimizes both the ellipsoidal shape and the mapping onto it, giving the highest flexibility among the three methods. Therefore, it gives the most uniformly distributed and regular triangle elements in the remeshing results.

For a more quantitative analysis, we evaluate the surface remeshing performance of different approaches by considering the shape and size variations of the triangle elements in the remeshed surfaces. Ideally, the triangle elements on the remeshed surface should be as uniform as possible in terms of their size (i.e. the area of the triangles) and shape (i.e. the regularity of the triangles). To assess the size variation of the remeshed surface, we define the size variation measure $\delta_{\text{size}}$ as follows: 
\begin{equation}
    \delta_{\text{size}} = \log \left(\frac{A_{\text{max}}}{A_{\text{min}}}\right),
\end{equation}
where $A_{\text{max}}$ is the maximum triangle area and $A_{\text{min}}$ is the minimum triangle area in the remeshed surface. It is easy to see that $\delta_{\text{size}} \geq 0$ for any remeshed surface, and the equality holds if and only if $A_{\text{max}} = A_{\text{min}}$, i.e. all triangle elements of the remeshed surface are equal in size. This shows that $\delta_{\text{size}}$ can effectively capture the size variation in the remeshed surface.

To assess the shape variation, we first define the face regularity $R_i$ for each face $T_i$ of the remeshed surface as follows:
\begin{equation}
R_i = \sum^3_{j = 1} \left| \frac{e^i_j}{e^i_1 + e^i_2 + e^i_3} - \frac{1}{3} \right|,
\end{equation}
where $e^i_1, e^i_2, e^i_3$ represent the length of the three edges of $T_i$. Note that $R_i = 0$ if and only if $T_i$ is an equilateral triangle. Then, we define the shape variation measure $\delta_{\text{shape}}$ of the entire remeshed surface as
\begin{equation}
    \delta_{\text{shape}} = \underset{i}{\text{mean}} (R_i).
\end{equation}
It is easy to see that $\delta_{\text{shape}} \geq 0$ for any remeshed surface, and the equality holds if and only if all triangles in the remeshed surface are equilateral. Therefore, $\delta_{\text{shape}}$ can effectively quantify the overall shape variation of the remeshed surface.

Table~\ref{tab:remshing_comparison} records the values of $\delta_{\text{size}}$ and $\delta_{\text{shape}}$ for all remeshing results shown in Fig.~\ref{fig:remeshing_pig}, Fig.~\ref{fig:remeshing_bear}, Fig.~\ref{fig:remeshing_hippocampus}, and Fig.~\ref{fig:remeshing_vaselion}. For the two conformal approaches (SCM and ECM), it can be observed that the values of $\delta_{\text{shape}}$ are small but the values of $\delta_{\text{size}}$ are large, which means the remeshing results obtained by those two methods are highly non-uniform in size. By contrast, the values of $\delta_{\text{size}}$ for the three density-equalizing approaches (SDEM, EDEM, and EDEQ) are small. This demonstrates the advantage of utilizing density-equalizing maps for surface remeshing. Furthermore, by considering both $\delta_{\text{size}}$ and $\delta_{\text{shape}}$ for the three density-equalizing methods, it can be observed that the EDEQ method gives the best overall remeshing performance among the three methods, which also matches our observations in Fig.~\ref{fig:remeshing_detail}.

\begin{table}[t]
\small
    \caption{\textbf{Comparison between the surface remeshing results obtained by SCM~\cite{choi2015flash}, ECM~\cite{choi2024fast}, SDEM~\cite{lyu2024spherical}, EDEM, and EDEQ.} For each method, we record the shape variation measure $\delta_{\text{shape}}$ and the size variation measure $\delta_{\text{size}}$ of the remeshed surface.}\label{tab:remshing_comparison}
    
  \begin{center}
  \begin{tabular}{|c|c|c|c|} \hline
    \bf Surface & \bf Method & \bf Shape variation $\delta_{\text{shape}}$  & \bf Size variation $\delta_{\text{size}}$  \\ \hline

    \multirow{5}{*}{Pig (Fig.~\ref{fig:remeshing_pig})} & SCM  & 0.0499 & 4.5846   \\ \cline{2-4}
     & ECM  & 0.0505  & 4.7123  \\ \cline{2-4}
    & SDEM  & 0.1562  &   3.3251   \\ \cline{2-4}
    & EDEM  & 0.1493  &  2.4706    \\ \cline{2-4}
    & EDEQ  & 0.1478 &   1.4218   \\ \hline
    
    \multirow{5}{*}{Bear (Fig.~\ref{fig:remeshing_bear})} & SCM  & 0.0535  &  3.7168  \\ \cline{2-4}
     & ECM  & 0.0526  & 3.9181  \\ \cline{2-4}
    & SDEM  & 0.1296  &  1.9349    \\ \cline{2-4}
    & EDEM  & 0.1282  &  1.4401   \\ \cline{2-4}
    & EDEQ  & 0.1263  &  1.3977   \\ \hline

    \multirow{5}{*}{Hippocampus (Fig.~\ref{fig:remeshing_hippocampus})} & SCM  & 0.0546  & 7.7204  \\ \cline{2-4}
     & ECM  & 0.0549  & 7.0796  \\ \cline{2-4}
     & SDEM  & 0.2593  & 4.1060 \\ \cline{2-4}
     & EDEM  & 0.2428  & 1.3482  \\ \cline{2-4}
     & EDEQ  & 0.1568  & 1.2508  \\ \hline
     
    \multirow{5}{*}{Lion-Vase (Fig.~\ref{fig:remeshing_vaselion})} & SCM  & 0.0587  & 5.5245  \\ \cline{2-4}
     & ECM  & 0.0576   & 5.4088   \\ \cline{2-4}
     & SDEM  & 0.1744 & 2.9125  \\ \cline{2-4}
     & EDEM  & 0.1583  & 1.7410  \\ \cline{2-4}
     & EDEQ  & 0.1249  & 1.7081   \\ \hline
  \end{tabular}
  
\end{center}
\end{table}

 \section{Discussion}\label{sect:discussion}
In this work, we have proposed a novel method for computing bijective ellipsoidal density-equalizing maps (EDEM) for genus-0 closed surfaces. Using this approach, we can efficiently obtain parameterizations with different desired mapping effects, including ellipsoidal area-preserving parameterizations and ellipsoidal parameterizations with controlled area change. Also, by considering a combined energy involving both a density-equalizing term and a quasi-conformal term, we can achieve ellipsoidal density-equalizing quasi-conformal maps (EDEQ) balancing the angle and area distortions. Experimental results on various ellipsoidal surfaces and genus-0 closed surface models have demonstrated the effectiveness of our proposed methods. 

Our proposed methods have broad applicability across various scientific and engineering domains. As shown in the previous section, our methods can be applied to surface remeshing for genus-0 closed surfaces with significantly improved mesh quality. Besides, the proposed methods can be applied to texture mapping in the field of computer graphics. In particular, since the proposed methods can reduce the area and angle distortions of the parameterization of genus-0 closed surfaces, high-quality texture mapping results can be effectively obtained. Also, as demonstrated by our experiments on the Hippocampus model, our proposed methods are well-suited for the analysis of complex anatomical structures in medical imaging. Moreover, analogous to the prior works on density-equalizing map projection, our EDEM and EDEQ methods can be naturally applied to cartogram creation, data visualization, and geophysical surface modeling.

We remark that while our proposed methods have considered both the density-equalizing property and the quasi-conformality, landmark-matching constraints have not been incorporated into our formulations. In the future, we plan to further extend our methods to facilitate landmark-matching surface parameterization and surface registration. Besides, note that some practical applications may involve curvature-based surface registrations, which use curvature to guide the registration process between surfaces. By combining curvature-based energies and our proposed density-based and quasi-conformal energies, it may be possible to extend our model to achieve curvature-based surface registrations with controlled area change for surfaces with extreme geometries. We also plan to perform more theoretical analyses on the convergence of our proposed algorithms in future work.

\bibliographystyle{ieeetr}
\bibliography{ellipsoidalDEMbib.bib}
\end{document}